# Formation of Hydrogen, Oxygen, and Hydrogen Peroxide in Electron Irradiated Crystalline Water Ice

Weijun Zheng,[1, 2] David Jewitt,[1] and Ralf I. Kaiser[2*]

[1] *Institute for Astronomy, University of Hawaii at Manoa, Honolulu, HI 96822*

[2] *Department of Chemistry, University of Hawaii at Manoa, Honolulu, HI 96822*

* Author to whom the correspondence should be sent.

Electronic mail: kaiser@gold.chem.hawaii.edu

**Abstract**

Water ice is abundant both astrophysically, for example in molecular clouds, and in planetary systems. The Kuiper belt objects, many satellites of the outer solar system, the nuclei of comets and some planetary rings are all known to be water-rich. Processing of water ice by energetic particles and ultraviolet photons plays an important role in astrochemistry. To explore the detailed nature of this processing, we have conducted a systematic laboratory study of the irradiation of crystalline water ice in an ultrahigh vacuum setup by energetic electrons holding a linear energy transfer of 4.3 ± 0.1 keV $\mu m^{-1}$. The irradiated samples were monitored during the experiment both on line and *in situ* via mass spectrometry (gas phase) and Fourier transform infrared spectroscopy (solid state). We observed the production of hydrogen and oxygen, both molecular and atomic, and of hydrogen peroxide. The likely reaction mechanisms responsible for these species are discussed. Additional formation routes were derived from the sublimation profiles of molecular hydrogen (90 – 140 K), molecular oxygen (147 – 151 K) and hydrogen peroxide (170 K). We also present evidence on the involvement of hydroxyl radicals and possibly oxygen atoms as building blocks to yield hydrogen peroxide at low temperatures (12 K) and via a diffusion-controlled mechanism in the warming up phase of the irradiated sample.

## 1. Introduction

Charged particle and ultra violet (UV) photon irradiation experiments of water ($H_2O$) ices at distinct temperatures are important means for an understanding of the chemical processing of the molecular clouds (Dressler 2001; Minh & Dishoeck 2000) and of the surfaces of outer solar system objects (Ehrenfreund et al. 1999; Sykes 2002). Water ice is not only the dominant component of ice mantles of dust grains in cold molecular clouds (Ehrenfreund, et al. 1999), but also the major constituent on the surfaces of small solar system objects such as Kuiper Belt Objects (KBOs) (de Bergh 2004; Jewitt & Luu 2004), comets (Campins, Swindle, & Kring 2004; De Pater & Lissauer 2001; Shematovich et al. 2005), and icy satellites like Europa, Ganymede, and Callisto (Taylor 2001) in the outer solar system. The icy dust grains in molecular clouds and on the surfaces of the solar system objects are subject to irradiation by energetic species from the solar wind (keV particles), the galactic cosmic radiation field (keV – PeV), UV photons, and potentially from particles trapped in planetary magnetospheres (Johnson 1990; Shematovich, et al. 2005). During the last quarter of a century, extensive laboratory experiments have been carried out to elucidate the radiation-induced formation of molecules in pure water ice targets; molecular hydrogen ($H_2$), molecular oxygen ($O_2$) and hydrogen peroxide ($H_2O_2$) have been proposed to be the most important irradiation products (Table 1).

Table 1. Previous experimental studies of irradiation effects on pure water ($H_2O$ or $D_2O$) ices

| | Reference | Thickness, nm | T, K | Particle | Energy, eV | Dose, eV molecule$^{-1}$ | Products | Pressure, torr |
|---|---|---|---|---|---|---|---|---|
| 1978 | (Brown et al. 1978) | 25-2500 | 15-110 | $He^+$ | $1.5×10^6$ | — | — | — |
| 1980 | (Brown et al. 1980) | 40-200 | 7-155 | $H^+$, $He^+$ | $0.9\text{-}1.5×10^6$ | — | — | — |
| 1980 | (Brown, Augustyniak, & Lanzerotti 1980) | 90-120 | 7-155 | $H^+$, $He^+$ | $6×10^3\text{-}1.8×10^6$ | — | — | — |
| 1984 | (Cooper & Tombrello 1984) | 190 | 10-60 | $F^{+q}$ | $1.6\text{-}2.5×10^6$ | — | — | $<10^{-9}$ |
| 1985 | (Bar-Nun et al. 1985) | — | 30-140 | $H^+$, $Ne^+$ | $0.5\text{-}6×10^3$ | — | $O_2$, $H_2$ and H | $<10^{-8}$ |
| 1986 | (Christiansen, Carpini, & Tsong 1986) | — | 78 | $Ar^+$, $Ne^+$, $N^+$, $He^+$, and $e^-$ | $2\text{-}6×10^3$ | — | — | — |
| 1993 | (Matich et al. 1993) | $(5.7±0.5)×10^5$ | 77 | UV | 4.77 | — | $O_2$ | $<5×10^{-7}$ |
| 1995 | (Kimmel & Orlando 1995) | 20 | 88 | $e^-$ | 5-120 | | $D(^2S)$, $O(^3P_{2,1,0})$, $O(^1D_{2,1})$ | $\sim1×10^{-10}$ |
| 1995 | (Shi et al. 1995a) | 150-250 | 20-120 | $O^+$ | $10\text{-}90×10^3$ | — | Mass loss | $10^{-10}$ -$5×10^{-9}$ |
| 1995 | (Shi et al. 1995b) | 120-150 | 60 | $H^+$, $D^+$, $He^+$, $Li^+$, $Be^+$, $B^+$, $C^+$, $N^+$, $O^+$, $F^+$, and $Ne^+$ | $10\text{-}100×10^3$ | — | Mass loss | $10^{-10}$ -$5×10^{-9}$ |
| 1996 | (Gerakines, Schutte, & Ehrenfreund 1996) | 100 | 10 | UV | >6 | — | $H_2O_2$, $HO_2$, and OH | $<8×10^{-8}$ |
| 1998 | (Sieger, Simpson, & Orlando 1998) | 15 | 90-150 | $e^-$ | <100 | <12 | $O_2$ | $\sim10^{-10}$ |
| 2000 | (Moore & Hudson 2000) | 3000<br>3000 | 16<br>80 | $H^+$<br>$H^+$ | $0.8×10^6$<br>$0.8×10^6$ | 0-17 | $H_2O_2$<br>No $H_2O_2$ | $10^{-7}$<br>$10^{-7}$ |
| 2001 | (Bahr et al. 2001) | — | 40-120 | $H^+$ | $200×10^3$ | — | $O_2$, $H_2O_2$ | $10^{-10}$ -$5×10^{-9}$ |
| 2003 | (Baragiola et al. 2003) | 150-250 | 20-120 | $H^+$<br>$He^+$<br>$O^+$<br>$Ar^+$ | $30×10^3$ | — | Mass loss | $\sim10^{-10}$ |
| 2003 | (Leto & Baratta 2003) | — | 16 | $H^+$<br>$Ar^{++}$<br>UV | $30×10^3$<br>$60×10^3$<br>10.2 | 0-100 | —<br>amorphization | $<8×10^{-8}$ |
| 2003 | (Orlando & Sieger 2003) | 15 | 110, | $e^-$ | 5-100 | — | $O_2$ | $2.0×10^{-10}$ |
| | | | 120 | | | | | |
| 2004 | (Gomis et al. 2004) | 500 | 16, 77 | $H^+$<br>$C^+$<br>$N^+$<br>$O^+$<br>$Ar^+$ | $30\times10^3$ | 0-200 | $H_2O_2$ | $<8\times10^{-8}$ |
| 2004 | (Gomis, Leto, & Strazzulla 2004) | 500 | 16, 77 | $H^+$<br>$He^+$<br>$Ar^{++}$ | $200\times10^3$<br>$200\times10^3$<br>$400\times10^3$ | 0-200 | $H_2O_2$ | $<8\times10^{-8}$ |
| 2004 | (Pan et al. 2004) | 2-3 | 87 | $e^-$ | 3-19 | — | $D_2O_2$, $DO_2$ | $\sim10^{-10}$ |
| 2005 | (Baragiola et al. 2005) | —<br>1900<br>373 | 100<br>80<br>70 | $e^-$<br>$H^+$<br>$Ar^+$ | $150\times10^3$<br>$100\times10^3$<br>$100\times10^3$ | — | $H_2O_2$ | — |
| 2005 | This work | 115 | 12 | $e^-$ | $5\times10^3$ | 0-500 | $H_2$, $O_2$, $H_2O_2$ | $5.0\times10^{-11}$ |

Considering molecular hydrogen, this species cannot be synthesized in the gas phase of molecular clouds via a simple recombination of two hydrogen atoms (Averna & Pirronello 1991; Pirronello & Averna 1988). Here, the reaction to form molecular hydrogen from the atoms is exoergic by about 4.5 eV; the hydrogen molecule formed in this process is highly rovibrationally excited and has to release this excess energy to avoid fragmentation via homolytic hydrogen-hydrogen bond rupture back to the atoms. Therefore, a third-body is required to transfer the excess energy. Due to the low number density in molecular clouds of only $10^4$ $cm^{-3}$, the collision frequency is too low to provide sufficient three-body collisions (Kaiser 2002); therefore, this pathway is not feasible under gas phase conditions in molecular clouds. Therefore, Flower & Pineau-Des-Forets (1990) postulated that molecular hydrogen is actually synthesized via recombination of two hydrogen atoms on the surfaces of grains in diffuse and dense molecular clouds. Also, molecular hydrogen can be formed by charged particle and UV photon processing of water and hydrogen-bearing molecules condensed on icy grains in cold molecular clouds. Sandford & Allamandola (1993) suggested that molecular hydrogen could be also trapped inside water-rich, porous icy grains even at temperatures as high as 12 K. Here, the trapped hydrogen might come from the adsorption of hydrogen onto dust grain surfaces, the trapping of $H_2$ inside the ice while the ice mantle evolves, or during *in situ* photolysis and radiolysis of the grain molecules – predominantly water - in the interstellar and also solar system ices (Sandford & Allamandola 1993; Dressler 2001; Minh & Dishoeck 2000; Sykes 2002; Ehrenfreund et al. 1999; de Pater & Lissauer 2001)

Molecular oxygen has been hypothesized to be one of the components of the icy interstellar dust grains (Greenberg, van de Bult, & Allamandola 1983). Based on the absorption bands at 577 nm and 627.5 nm, solid molecular oxygen has also been identified on the surfaces of the Jovian satellites Ganymede (Spencer, Calvin, & Person 1995) as well as on Europa and Callisto (Spencer & Calvin 2002). The presence of molecular oxygen in the atmospheres of Europa and Ganymede has also been reported. The far-ultraviolet oxygen airglow measurement inferred vertical column densities of molecular oxygen in the range $2.4 – 14 \times 10^{14}$ $cm^{-2}$ for Europa and $1 – 10 \times 10^{14}$ $cm^{-2}$ for Ganymede (Hall et al. 1998; Hall et al. 1995). Based on the detectable abundances of ozone ($O_3$), it has been further proposed that Saturn's satellites Rhea and Dione also have a sustainable amount of molecular oxygen in their atmospheres (Noll et al. 1997). Here, the molecular oxygen in the surfaces and atmospheres of these satellites could be formed from the radiolysis and energetic particles sputtering of water ice by energetic electrons ($e^-$), protons ($H^+$), and oxygen ions ($O^+$) trapped in Saturn's magnetosphere (Noll, et al. 1997).

Water-bonded hydrogen peroxide molecules have been suggested tentatively to be the carrier of the low frequency wing of the 3250 $cm^{-1}$ infrared absorption feature of the dust grains in cold molecular clouds (Tielens & Hagen 1982). Absorption features of hydrogen peroxide have been identified in the infrared and ultraviolet wavelength spectra of Europa, where the hydrogen peroxide concentration is about 0.13 ± 0.07% relative to water (Carlson et al. 1999). Hendrix et al. (1999) reported the existence of hydrogen peroxide on the surfaces of Europa, Ganymede, and Callisto with concentrations of about 0.3 % by weight. The authors speculate that hydrogen peroxide probably originates from an irradiation of water ice by energetic particles in those systems. Note that $H_2O_2$ has also been proposed as a precursor molecule to generate molecular oxygen on the surfaces of Ganymede and Europa (Sieger, Simpson, & Orlando 1998).

To understand the radiation induced chemistry in water ice, systematic laboratory experiments have to be carried out. These studies are aimed to mimic the chemical composition, temperature, and radiation environment of molecular clouds and of outer solar system objects (Johnson & Quickenden 1997). The laboratory experiments should identify the molecules generated inside water ice both qualitatively and quantitatively. Finally, a major objective is to determine the reaction mechanisms and the chemical dynamics of the processes involved; this will help us to predict the formation of complex, oxygen bearing molecules in the interstellar medium and in our solar system (Kaiser 2002).

The formation of molecular hydrogen on dust grains has been investigated both theoretically and experimentally. Averna & Pirronello (1991; Pirronello & Averna 1988) conducted Monte Carlo simulations to tackle the formation of molecular hydrogen via cosmic ray particles irradiation of dust grain mantles in dense clouds. Brown et al. (1982) performed a quantitative study of molecular hydrogen ($H_2$) and deuterium ($D_2$) sputtered from water ices irradiated by high-energy (keV-MeV) protons as a function of temperature. Bar-Nun et al. (1985) studied the ejection of $H_2O$, $O_2$, $H_2$ and H from water ice at 30 – 140 K bombarded by 0.5 – 6 keV $H^+$ and $Ne^+$. Kouchi & Kuroda (1990) reported the thermal desorption spectra of $H_2$ molecules from the UV-irradiated crystalline (cubic) water ice. Sandford & Allamandola (1993) investigated the formation of $H_2$ molecules in UV photolysis of ice mixtures, whereas Kimmel et al. (1996; 1994) measured the electron-stimulated $D_2$ ($H_2$) yield desorbed directly from amorphous ice at 50 K by 5 – 50 eV electron irradiation. Westley et al. (1995) observed the desorbing molecular hydrogen by irradiating water ice with 121.6 nm (Lyman-ALPHA) photons at 50 – 100 K; on the other hand, Watanabe et al. (2000) measured the yield of $D_2$ via ultraviolet irradiation of amorphous heavy water ($D_2O$) ice 12K. Note that the formation of molecular hydrogen and deuterium in amorphous water has also been studied by Manico et al. (2001), Roser et al. (2002), and Hornekaer et al. (2003; 2005).

Considering molecular oxygen, the studies are relatively sparse. This is due to the fact that an individual oxygen molecule is infrared inactive, and hence, difficult to detect via commonly used infrared spectroscopy in laboratory experiments. Matich et al.(1993) studied the $O_2$ luminescence from $H_2O$ and $D_2O$ ices excited with 260 nm photons at low temperatures. Vidal et al. (1997) conducted laboratory experiments to investigate the origin of oxygen absorption bands in the visible reflectance spectrum of Ganymede. Sieger et al. (1998) probed the production of $O_2$ in $D_2O$ ice by electron-beam irradiation relating their findings qualitatively to Ganymede and Europa. These authors proposed a two-step mechanism for the formation of molecular oxygen via precursor molecules in low-temperature water ice. Orlando and Sieger (2003) measured the electron energy threshold, fluence, and temperature dependence for $O_2$ production during low-energy (5–100 eV) electron bombardment of amorphous and crystalline $D_2O$ ice films.

The formation of the hydrogen peroxide molecule also received much attention. Gerakines et al. (1996) investigated the ultraviolet processing of interstellar ice analogs. Hydrogen peroxide was identified via infrared absorption bands at 2850 $cm^{-1}$ and 1457 $cm^{-1}$. Moore and Hudson (2000) carried out infrared spectroscopy studies of the formation of hydrogen peroxide in water ice irradiated with 0.8 MeV protons at temperatures relevant to the icy Galilean satellites (80–150 K). Gomis et al. (2004a; 2004b) performed an extensive infrared spectroscopy study of water ice exposed to $H^+$, $He^+$, $Ar^{++}$, $C^+$, $N^+$, $O^+$ and $Ar^+$ charged particles. They found that the produced $H_2O_2/H_2O$ ratio is greater for the heaver ions (~6% for the case of $Ar^{++}$, only ~1% for $H^+$). Bahr et al. (2001) conducted quantitative laboratory radiolysis experiments on cubic water ice between 40 and 120 K with 200 keV protons. They found that trapped $O_2$ and $H_2O_2$ can be released from the irradiated ice upon warming. Pan et al. (2004) suggested that the production of hydrogen peroxide may be initiated by the dissociative electron attachment (DEA) of low-energy electrons (LEEs) to water molecules followed by recombination of two hydroxyl (OH) radicals. Electronic excitation or ionization by electrons also contributes to the formation of hydrogen peroxide at higher electron energies. The authors also proposed that the hydroperoxyl ($HO_2$) radical could be formed on the surfaces of icy satellites by low energy electron impact. Finally, Baragiola et al. (2005) studied the irradiation of water ice by 150 eV electrons and 100 keV ions using low-energy electron-energy-loss spectroscopy (EELS) and infrared spectroscopy.

Despite these extensive investigations of the radiation chemistry of water (Table 1), a quantitative understanding of the formation mechanisms and the reaction dynamics of molecular hydrogen and oxygen as well as of hydrogen peroxide is far from being accomplished. This knowledge is crucial to quantitatively predict the chemical and astrobiological evolution of distinct extraterrestrial environments. Recently, we established a systematic research program to untangle the charged particle and photon induced high energy chemistry in water ices. The goal is to investigate the formation of molecular and atomic hydrogen and oxygen as well as hydrogen peroxide together with ozone quantitatively, as a function of important astrophysical parameters. These are: i) the irradiation dose absorbed by a water molecule (energy deposition into the water ice sample), ii) the temperature (~10 K (molecular clouds; comets in Oort cloud) to ~150 K (Jovian satellites)), iii) the modification of the water ice (crystalline versus amorphous), iv) the irradiating particle (electron, proton, photon) and their energies and wavelengths, respectively. To collect quantitative data, these laboratory simulation experiments are carried out in a single machine under ultra high vacuum (UHV) conditions by monitoring the complete product spectrum of the newly formed species.

In this paper, we report a systematic study of the irradiation of crystalline water ice with energetic electrons at 12 K to obtain a quantitative understanding of the formation mechanisms and the reaction dynamics of molecular hydrogen and oxygen as well as of hydrogen peroxide in crystalline ices. It is important for each laboratory astro chemist to carry out first experiments under well defined conditions (here: crystalline water ice) before going to a more complex system (here: amorphous water ice). This helps to better understand the reaction mechanisms in electron – irradiated samples. - The choice of the low temperature characterizes the condition of water ice in cold molecular clouds and on comets located in the Oort cloud. Electrons were chosen to mimic the secondary electrons released via ionization processes in the track of cosmic ray particles and energetic electrons in the magnetospheres of planets. Recall that 99.99 % of the kinetic energy transferred to the target molecules by the galactic cosmic ray particles is released via the electronic interaction potential leading to electronic excitation, vibrational excitation, bond ruptures, and ionization (Bennett et al. 2004; Jamieson et al. 2005; Kaiser 2002); therefore, electron irradiation experiments present the simplest way elucidate whether electronic and/or nuclear interaction with target molecules/atoms is the driving force to synthesize new molecules in water rich ices. Also, in the laboratory experiment, even at ultra high vacuum of $5 \times 10^{-11}$ torr, amorphous water ice can usually trap residual gases – mostly molecular hydrogen from the stainless steel walls of the reaction vessel. These trapped species make the chemical processing of low temperature ices more complicated. Therefore, we chose to investigate the chemical processing of crystalline water ice first in order to simplify the problem. Finally, in the present study we used both Fourier Transform Infrared Spectroscopy (FTIR) and quadrupole mass spectrometry (QMS) to detect the infrared active and infrared inactive species simultaneously in one experiment. Note that none of the previous investigations on water ice (Table 1) detected all species ($H_2$, $O_2$ and $H_2O_2$ together with O and H) simultaneously. Since the chemical dynamics and formation mechanisms of these species are closely inter-related, a synchronized detection of all atomic and molecular species, as done in the present experiments, is imperative. This allows us to examine the correlation between those products and obtain more accurate information about the mechanisms.

## 2. Experimental

The experiments were carried out in a contamination-free ultrahigh vacuum chamber (Bennett, et al. 2004). The top view of this apparatus is shown in Figure 1. Briefly, the setup consists of two units: the main chamber and the differentially pumped charged particle module (electron source). The main chamber can be evacuated by a magnetically levitated turbo molecular pump (1100 $ls^{-1}$) down to $5 \times 10^{-11}$ torr; the differential regions are pumped down to the low $10^{-10}$ torr regime by two 400 $ls^{-1}$ maglev pumps. All turbo molecular pumps are backed by oil-free scroll pumps. A two stage closed cycle helium refrigerator coupled with a differentially pumped rotary platform is attached to the main chamber and holds a polished polycrystalline silver mirror; the latter serves as a substrate for the ice condensation. With the combination of the closed cycle helium refrigerator and a programmable temperature controller, the temperature of the silver mirror can be regulated precisely (± 0.3 K) between 10 K and 350 K. Most important, the oil-free, ultra high vacuum pumping scheme guarantees that even at temperatures as low as 10 K, it takes about 200 h for one monolayer of the residual gases to condense on the silver mirror.

The water ices were prepared by condensing water vapor onto the silver substrate at 140 K. This temperature ensures the formation of cubic crystalline water ice (Jenniskens, Blake, & Kouchi 1998). To minimize the contamination from air inside the water ice, we froze triply distilled water with liquid nitrogen and defrosted it in vacuum for several times. The gas reservoir was pumped down to $10^{-7}$ torr before it was filled with about 11 torr water vapor. During the deposition, the water vapor pressure in the main chamber was maintained at $6 \times 10^{-9}$ torr for 9 minutes. To retain the cubic structure of the water ice, we cooled down the sample slowly from 140 K to 12 K with a cooling rate of 1.0 K $min^{-1}$. Note that the absolute thickness of the sample can be controlled reproducibly by the condensation time and the condensation pressure inside the main chamber. Figure 2 (solid line) depicts a typical infrared spectrum of the crystalline water ice sample recorded prior to the irradiation at 12 K; the absorptions are compiled in Table 2. Please note that our spectrum was recorded in an absorption-reflection-absorption mode. Therefore, the laboratory spectra – due to the detectable longitudinal and transversal modes – are different from astronomically recorded spectra and cannot be compared directly with the latter. However, it is not our prime interest to provide laboratory spectra which can be compared to astronomical ones in this work; this paper focuses on the reaction dynamics and mechanisms on the electron irradiation of water ices.

To determine the ice thickness quantitatively, we integrated the infrared absorption features at 3280 $cm^{-1}$, 1660 $cm^{-1}$, and 760 $cm^{-1}$ and calculated the column density $N$ (molecules $cm^{-2}$) via a modified Lambert-Beers relationship via equation (1) (Bennett, et al. 2004).

$$N = \frac{\ln 10}{2} \times \cos\alpha \times \frac{\int_{v_1}^{v_2} A(v)dv}{A_{ref}} \quad (1)$$

Here, the division by a factor of two corrects for the ingoing and outgoing infrared beam, α is the angle (75°) between the surface normal of the silver mirror and the infrared beam, $A_{ref}$ is the literature value of the integral absorption coefficient (cm $molecule^{-1}$), and $\int_{v_1}^{v_2} A(v)dv$ is the integral of the infrared absorption feature for our sample ($cm^{-1}$). Considering the integrated absorption coefficients of these fundamentals (Table 3), this treatment suggests column densities of 3.6 ± 0.9 × $10^{17}$ molecules $cm^{-2}$. Accounting for the density of crystalline water ice of 0.93 ± 0.02 g $cm^{-3}$ (Jenniskens, et al. 1998) and the molecular mass (18 g $mol^{-1}$), this translates into a thickness of 115 ± 30 nm. These ices were irradiated then at 12 K with 5 keV electrons for 180 min at nominal beam currents of 0 nA (blank experiment), 10 nA, 100 nA, 1000 nA, and 10000 nA by scanning the electron beam over an area of 1.86 ± 0.02 $cm^2$. Note that the actual extraction efficiency of the electron gun is 92 ± 4 %. After each irradiation, the sample was kept at 12 K for 60 minutes and then warmed up at 0.5 K per minute to 293 K.

To guarantee an identification of the reaction products in the ices and those subliming into the gas phase on line and *in situ*, a Fourier transform infrared spectrometer (FTIR; solid state) and a quadrupole mass spectrometer (QMS; gas phase) were utilized. The Nicolet 510 DX FTIR spectrometer (6000-500 $cm^{-1}$) operated in an absorption–reflection–absorption mode (resolution 2 $cm^{-1}$; each spectrum had been averaged for 120 sec). The infrared beam passed through a differentially pumped potassium bromide (KBr) window, was attenuated in the ice sample prior to and after reflection at a polished silver wafer, and exited the main chamber through a second differentially pumped KBr window before being monitored via a liquid nitrogen cooled detector (MCT-B type). The gas phase was monitored by a quadrupole mass spectrometer (Balzer QMG 420; 1–200 amu mass range) with electron impact ionization of the neutral molecules in the residual gas analyzer mode at electron energies of 100 eV.

Table 2. The infrared absorption bands of crystalline water ice at 12 K (Hagen, Tielens, & Greenberg 1981; Jenniskens et al. 1997).

| absorption, $cm^{-1}$ (μm) | assignment | characterization |
|---|---|---|
| 941 (10.6) | $v_L$ | libration band |
| 1574 (6.3) | $v_2$ | bending |
| 2270 (4.4) | $v_{L+}v_2$ / $3v_L$ | combination band / libration overtone |
| 3107 (3.2) | $v_1$ (in phase) | symmetric stretch |
| 3151 (3.3) | $v_3$ | asymmetric stretch (transversal mode) |
| 3332 (3.0) | $v_3$ | asymmetric stretch (longitudinal mode) |
| 3452 (2.9) | $v_1$ (out of phase) | symmetric stretch |
| 4900 (2.0) | $v_2+v_3$ or $v_2+v_1$ | combination band |

Table 3. The thickness of the water ice samples calculated for different absorption features. Absorption coefficients were taken from Gerakines et al. (1995).

| peak, $cm^{-1}$ | peak area, $cm^{-1}$ | integral absorption coefficient $A_{ref}$, cm $molecule^{-1}$ | column density N, molecule $cm^{-2}$ | thickness d, nm |
|---|---|---|---|---|
| 3683 – 2887 | 161.6 | $2.0\times10^{-16}$ | $2.4\times10^{17}$ | 77 |
| 1970 – 1245 | 18.2 | $1.2\times10^{-17}$ | $4.5\times10^{17}$ | 144 |
| 1078 – 567 | 40.5 | $3.1\times10^{-17}$ | $3.9\times10^{17}$ | 124 |

## 3. Results

### 3.1. Infrared Spectroscopy

The analysis of the infrared spectra is carried out in three consecutive steps. First, we investigate the newly emerging absorption features qualitatively and assign their carriers. Next, the variations of these absorptions in time upon electron irradiation are investigated quantitatively. Finally, these data are fit to calculate production rates of synthesized molecules in units of molecules $cm^{-2}$ (column density). The effects of the electron bombardment of the water ice samples are compiled in Figures 2 and 3. A comparison of the pristine sample with the exposed ices at 12 K clearly identifies novel absorption features (Figure 2 and Table 2). Even at the highest electron current of 10,000 nA, only the most prominent absorption feature of hydrogen peroxide ($H_2O_2$) emerged at 2851 ± 2 $cm^{-1}$ (3.5 μm) (Table 4). This band has a full width at half maximum of about 45 $cm^{-1}$. The intensity of hydrogen peroxide absorption at 2851 $cm^{-1}$ increases with increasing electron current. This band had also been observed by a number of groups during UV photolysis or energetic particle irradiation of water ice, i.e. (Gerakines, Schutte, & Ehrenfreund 1996), (Moore & Hudson 2000), and (Gomis, et al. 2004a). The assignment of this band will be discussed in the next paragraph. In Figure 2 (dash line), the infrared absorption features at region 3000-3700 $cm^{-1}$ after irradiation are slightly different from those before the irradiation. That may due to the amorphization of the water ice by electron irradiation. The absorption bands of the hydrogen peroxide might also contribute there since hydrogen peroxide has $v_5$ asymmetric stretch at 3175 $cm^{-1}$ and $v_1$ symmetric stretch at 3285 $cm^{-1}$ (Giguere & Harvey 1959). However, we cannot resolve the hydrogen peroxide bands at that region due to the high intensities of the water bands and the low concentration of the hydrogen peroxide (see quantitative results below). Absorption features of the hydroperoxyl radical ($HO_2$) and ozone ($O_3$) were not detected.

During the warming up phase two additional absorption features of the hydrogen peroxide emerged at 170 K, i.e. after most of the water sublimed into the gas phase (Figure 3). These peaks are centered at 2848 $cm^{-1}$, 1454 $cm^{-1}$, and 1389 $cm^{-1}$, respectively. The remaining absorption bands belong to water ice (Table 2). The 1454 $cm^{-1}$ mode was identified as $v_2$ symmetric bending (Giguere & Harvey 1959), whereas the 1389 $cm^{-1}$ absorption presents the $v_6$ asymmetric bending (Giguere & Harvey 1959). The 2848 $cm^{-1}$ peak is slightly red-shifted compared to the 12 K sample owing to the difference of temperature. Its frequency is approximately the sum of the peaks at 1454 $cm^{-1}$ and 1389 $cm^{-1}$. Therefore, we assign it as the $v_2+v_6$ combination mode which is consistent with previous experimental studies (Moore & Hudson 2000).

In order to investigate the yield of hydrogen peroxide ($H_2O_2$) and its relationship to the irradiation time and hence the dose, we integrated the $v_2+v_6$ combination band in each spectrum; utilizing the integral absorption coefficient of ($A_{(H2O2)} = 2.7\times10^{-17}$ cm molecule$^{-1}$, Moore et al. 2002) we can apply equation (1) to compute the column density. Figure 4 shows the change of hydrogen peroxide column density versus the irradiation time. It is important to stress that even the strongest $v_2+v_6$ absorption overlaps with low frequency side of the $v_1$ stretching of the water molecule. As a matter of fact, the overlap between the water feature and hydrogen peroxide feature is so significant that at a nominal irradiation current of 10 nA, the $\nu_2+\nu_6$ absorption does not 'emerge' from this shoulder; similarly, a solid identification in the 100 nA irradiation experiments is only feasible after 70 min. On the other hand, in both the 1000 nA and 10000 nA experiments, the 2848 $cm^{-1}$ band is visible right after the onset of the irradiation. Comparing the initial column density of the water with the hydrogen peroxide after the irradiation exposure (180 minutes) suggests that 0.4 %, 1.3 %, and 3.3 % of the water were converted to hydrogen peroxide at irradiation currents of 100 nA, 1000 nA, and 10000 nA, respectively. We would like to stress again that these yields are obtained at 12 K; also, the peaks at 10 nA currents were too weak to be detected. To obtain information on possible temperature-dependent formation mechanisms of hydrogen peroxide, we warmed up the irradiated sample until most of the water sublimed into the gas phase (Figure 3). Finally, it is important to point out that no infrared absorption feature of molecular hydrogen ($H_2$)(4143 $cm^{-1}$) or molecular oxygen ($O_2$)(1549 $cm^{-1}$) was observed since those molecules are infrared-inactive. Therefore, their quantities cannot be determined from the infrared spectra, and mass spectrometric data are crucial.

Table 4. Observed infrared absorption bands of hydrogen peroxide ($H_2O_2$).

| absorption, $cm^{-1}$ (μm) | assignment | characterization |
|---|---|---|
| 2848 (3.5) | $\nu_2+\nu_6$ | combination band |
| 1454 (6.9) | $\nu_2$ | symmetric bending |
| 1389 (7.2) | $\nu_6$ | asymmetric bending |

### 3.2. Mass Spectrometry

We are now comparing the infrared observations with a mass spectrometric analysis of the gas phase. During the irradiation phase of the sample, signals at m/e = 2 ($H_2^+$) and m/e = 1 ($H^+$) were observed; the temporal development of the molecular hydrogen ion – converted to partial pressure - is shown in Figure 5. It is evident that as the irradiation started, the partial pressure of molecular hydrogen rose sharply at irradiations with 1000 nA and 10000 nA; at 10 nA and 100 nA, the signal was too weak to be detected. Once the electron irradiation has stopped, the signal at m/e = 2 decreased, too. This suggests that the molecular hydrogen originates from the electron bombardment; also, blank experiments showed that the molecular hydrogen is not an artifact from the residual gas (Figure 5). It is important to address the temporal evolution of the ion currents of m/e = 2 and m/e = 1 further. A quantitative calibration of our mass spectrometer (Appendix A) showed that neutral molecular hydrogen molecules are ionized to $H_2^+$ (m/e = 2), but also fragment to H + $H^+$ (m/e = 1) in the electron impact ionizer; quantitatively spoken, the ratio of the ion currents at m/e = 1 to m/e = 2 was determined to be $7.3 \pm 0.7 \times 10^{-3}$. During the irradiation experiments at 1000 nA and 10000 nA, we detected not only signal at m/e = 2, but also from $H^+$ at m/e = 1 (the ion current at m/e = 1 at the 10 nA and 100 nA experiments was too small to be detected). Here, ratios of $H^+/H_2^+ = 2.2 \pm 0.3 \times 10^{-2}$ (1000 nA) and $7.4 \pm 0.7 \times 10^{-2}$ (10000 nA) were derived. The enhanced ion currents of $H^+$ by factors of about three and ten compared to the fragmentation pattern of the neat hydrogen gas strongly suggest that not only molecular hydrogen, but also hydrogen atoms are released during the electron exposure of the water ice into the gas phase. Accounting for the ionization cross sections of 0.55 $A^2$ (H) and 0.97 $A^2$ ($H_2$) and of the enhanced pumping speed of molecular hydrogen atoms by a factor of about 1.4 (Chambers, Fitch, & Halliday 1998), we derive relative production rates of atomic hydrogen of 0.6 ± 0.2 % and 2.6 ± 0.4 % compared to molecular hydrogen released into the gas phase *during* the irradiation experiments at 1000 nA and 10000 nA, respectively. Finally, we also observed an ion current at m/e = 16 ($O^+$) with the start of the electron exposure in both experiments. Considering the ionization cross section of atomic oxygen of 1.4 $A^2$ and approximating the pumping speed to be a factor of two larger than molecular hydrogen (Chambers, et al. 1998), we derive relative formation rates of 35 ± 8 % and 45 ± 14 % of atomic oxygen relative to molecular hydrogen.

After the electron exposure, the samples were kept isothermally at 12 K and were then warmed up with heating rate of 0.5 K $min^{-1}$. This releases newly formed molecules into the gas phase. The reader should keep in mind that the successive thermal processing of the irradiated sample induces thermal, mostly diffusion controlled reactions. This changes the chemical composition of the irradiated ice sample. Therefore, a careful correlation of the mass spectra with the infrared spectra during the warm up phase is required (see section 4). Figure 6 depicts the ion currents of molecular hydrogen (m/e = 2; $H_2^+$), molecular oxygen (m/e = 32; $O_2^+$), water (m/e = 18; $H_2O^+$), and hydrogen peroxide (m/e = 34; $H_2O_2^+$) during the warm up phases of the blank run (a) and the irradiation experiments at 10 nA (b), 100 nA (c), 1000 nA (d), and 10000 nA (e) versus the temperature. Most importantly, we did not detect any ozone ($O_3$) during the warm up phase. This suggests that the partial pressure of ozone formed during the electron exposure is less than $10^{-15}$ torr (the detection limit of our mass spectrometer). Due to this low detection limit, it is crucial to comment on the blank experiment (Figure 6 (a)). Here, we observe a small peak of molecular hydrogen during the warm up of the sample peaking at 19 K. Note that even in an extreme ultra high vacuum chamber at $5 \times 10^{-11}$ torr, molecular hydrogen residual gas can never be eliminated quantitatively due to the outgassing of the stainless steel chamber. The peak at 19 K probably resulted from hydrogen molecules adsorbed on the ice surface after the sample has been cooled to 12 K. To test this hypothesis, we conducted another kind of blank experiment: cooling the silver substrate to 12 K without condensing water ice. Here, we also detected a molecular hydrogen peak centered at about 17 K strongly suggesting that the molecular hydrogen from the residual gas also condenses on the silver wafer. We also irradiated the neat silver wafer, i.e. without water ice; the results were the same as the unirradiated neat silver wafer. Note that in the experiment without water ice, molecular hydrogen starts subliming at a slightly lower temperature compared to blank experiment of the water sample; this may be because of a higher bonding energy of molecular hydrogen on the water surface compared to the silver wafer. As the temperature reaches 140 K, water starts to sublime; the broad peaks of the m/e = 2 and m/e = 32 ($H_2^+$ and $O_2^+$, respectively) between 140 - 170K originate from the residual $H_2$ and $O_2$ trapped inside the ice and the fragmentation of the water molecule in the electron impact ionizer of the mass spectrometric detector (Figure 6 (a)). Most importantly, no hydrogen peroxide is visible in the blank runs.

We next discuss the mass spectra obtained during the warm up phases (Figure 6 (b)-(e)). Comparing the blank sample (Figure 6 (a)) with the irradiation experiments clearly shows that the sharp molecular hydrogen peaks at about 19 K originate from surface contamination, as discussed above. On the other hand, the broad hydrogen peaks starting at 114 K (b) and 90 K (c – e) cannot be seen in the blank spectrum. Therefore, these patterns are a direct proof that molecular hydrogen is released during the warm up phase as a result of the electron bombardment. Recall that in each irradiation experiment, a third molecular hydrogen peak appears during the sublimation of the water sample similar to the blank experiment. Besides the molecular hydrogen, we were also able to monitor the newly formed hydrogen peroxide during the warm up phases at m/e = 34; $H_2O_2^+$. Here, hydrogen peroxide was released at temperature region 160 - 180 K with the center around 170K. Also, comparing the signal with the blank run (Figure 6 (a)) verifies that hydrogen peroxide is not a contaminant of our system, but clearly a product of the electron exposure. Finally, we also observe molecular oxygen (m/e = 32) peaking at about 147 – 151 K (100 nA – 10000 nA); the oxygen peak can be seen more easily by comparison of Figure 6 (a) and Figure 6 (d). Note that no oxygen from irradiation could be detected at the 10 nA irradiation experiments except the broad oxygen peak similar to the blank experiment.

Similarly to the infrared data, we attempt now to quantify the mass spectrometric data and extract formation rates and absolute yields of the newly formed molecules. This requires a detailed calibration of the mass spectrometer and of the ion gauges utilized in the setup (Appendix A). Here, we integrated the ion currents of the newly formed molecular hydrogen, molecular oxygen, and hydrogen peroxide species; these total ion currents can be converted to the total number of synthesized molecules (Appendix A). The results are summarized in Table 6 and in Figure 7. As the electron current of the experiment increase from 10 nA to 10000 nA, the absolute number of hydrogen, oxygen, and hydrogen peroxide molecules rises sharply (10 – 1000 nA) and then saturates quickly once the current reaches 10000 nA. Note the irradiation at 10 nA and 10000 nA for 180 minutes correspond to $3.33\times 10^{14}$ and $3.33\times 10^{17}$ electrons/cm$^2$ in our experiments. Qualitatively, these patterns are similar for all three molecular species observed (Figure 7 (a) – (c)). Also, it is worth noting from Figure 7 that molecular hydrogen is the dominant reaction product followed by hydrogen peroxide and molecular oxygen. These data can be utilized now to extract information about the reaction mechanisms of the newly formed species.

## 4. Discussion

### 4.1. Formation of Hydrogen Peroxide

To investigate the response on crystalline water at 12 K on energetic electrons quantitatively, we must have a close look at the temporal evolution of the infrared spectroscopic (Figure 4) and mass spectrometric data (Figure 7; Table 6). The evolution of the column density of the hydro-gen peroxide ($H_2O_2$) molecule during the irradiation (Figure 4) can be fit successfully with a (pseudo) first order reaction following the kinetics via equation (2) utilizing the following parameter a(100 nA) = $2.2 \pm 0.7 \times 10^{15}$ $cm^{-2}$, a(1000 nA) = $2.7 \pm 0.8 \times 10^{15}$ $cm^{-2}$, a(10000 nA) = $6.0 \pm 1.8 \times 10^{15}$ $cm^{-2}$, k(100 nA) = $0.003 \pm 0.001$ $min^{-1}$, k(1000 nA) = $0.06 \pm 0.02$ $min^{-1}$, and k(10000 nA) = $0.16 \pm 0.05$ $min^{-1}$.

(2) $$[H_2O_2](t) = a\,(1 - e^{-k\,t}).$$

We would like to comment briefly on the functional form of this fit. In chemical kinetics, equation (2) dictates that a chemical reaction forming hydrogen peroxide proceeds formally via a unimolecular decomposition (here: an electron induced decomposition) of a precursor molecule/ complex *A* via first order kinetics in a one-step mechanism holding a rate constant k (equation (3)); X defines potential decomposition by-products. However, a chemical reaction can also occur from a reactant molecule/complex *A* via an intermediate to form then a hydrogen peroxide molecule (equation (4)). If the rate constant of the first step from *A* to *B* is much faster than from *B* to hydrogen peroxide, this reaction is said to proceed via *pseudo* first order. Both reaction sequences (3) and (4) can account for the formation of hydrogen peroxide via equation (2). The infrared spectroscopic data suggest that at the end of the irradiation, about $0.44 \pm 0.15$ % (100 nA), $1.3 \pm 0.4$ % (1000 nA), and $3.3 \pm 1.0$ % (10000 nA) of the water molecules are being converted to hydrogen peroxide.

(3) $$A \rightarrow H_2O_2 + X.$$

(4) $$A \rightarrow B \rightarrow H_2O_2 + X.$$

These data help us to unravel the nature of the initial complex *A* which decomposes upon electron irradiation via equations (3) and/or (4). Two reaction sequences are feasible. First, comparing the electronic structure of the water and the hydrogen peroxide molecules, it is necessary to cleave at least one oxygen-hydrogen bond of the water molecule forming the hydroxyl radical ($OH(X^2\Pi_\Omega)$) plus atomic hydrogen (equation (5)). This process is endoergic by 466.1 $kJmol^{-1}$. This pathway has also been proposed by previous works (Gomis, et al. 2004a; Gomis, et al. 2004b; Pan, et al. 2004). Two of the hydroxyl radicals can recombine in an exoergic reaction (($\Delta_R G$= -174.0 $kJmol^{-1}$) without barrier to form a hydrogen peroxide molecule (equation (6)). Note, however, that at 12 K, the hydroxyl radical is not mobile. Therefore, reaction (6) only occurs if both hydroxyl radicals are born in *neighboring* sites within the ice matrix; also, both ra-dicals must hold the correct *recombination geometry*. Secondly, recent crossed molecular beam data of the reaction of electronically excited oxygen atoms ($O(^1D)$) with molecular hydrogen ($H_2(X^1\Sigma_g^+)$) (Liu 2001) suggest that one micro channel follows indirect scattering dynamics via an insertion of the oxygen atom into the hydrogen-hydrogen bond of the hydrogen molecule thus forming a water molecule. The microscopic reversibility dictates that upon internal excitation of the water molecule (vibrational energy), the latter can decompose in one step to form electronically excited oxygen atoms plus molecular hydrogen (reaction (7)). This reaction requires 650.1 $kJmol^{-1}$ energy to be transferred from the energetic electron to the water molecule. The electronically excited oxygen atom can react now with a *neighboring* water molecule via a barrier-less insertion into an oxygen-hydrogen bond to form hydrogen peroxide (equation (8)) (Ge, Head, & Kaiser 2005). Alternatively, the oxygen atom can add barrier-lessly to the oxygen atom of the water molecule forming a short-lived oxywater intermediate which rearranges then via hydrogen migration to the hydrogen peroxide molecule (equation (9)). The overall reaction exothermicities of reactions (8) and (9) were determined to be 298.5 $kJmol^{-1}$. Therefore, to fit the temporal profiles of the hydrogen peroxide column densities (Figure 4) at 12 K, equations (5) and (6) and (7) – (9) can be recombined and rewritten as equations (10) and (11), respectively. In both cases, these equations suggest an electron induced decomposition of two neighboring water molecules $[(H_2O(X^1A_1))_2]$ to form the hydrogen peroxide molecule at 12 K. The relative contribution of equation (10) versus (11) is still under investigation in our lab.

(5) $$H_2O(X^1A_1) \rightarrow H(^2S_{1/2}) + OH(X^2\Pi_\Omega)$$

(6) $$2\ OH(X^2\Pi_\Omega) \rightarrow H_2O_2(X^1A)$$

(7) $$H_2O(X^1A_1) \rightarrow O(^1D) + H_2(X^1\Sigma_g^+)$$

(8) $$O(^1D) + H_2O(X^1A_1) \rightarrow H_2O_2(X^1A)$$

(9) $$O(^1D) + H_2O(X^1A_1) \rightarrow [OOH_2(X^1A)] \rightarrow H_2O_2(X^1A)$$

(10) $$[(H_2O(X^1A_1))_2] \rightarrow [H(^2S_{1/2})\ldots HO(X^2\Pi_\Omega)\ldots OH(X^2\Pi_\Omega)\ldots H(^2S_{1/2})] \rightarrow H_2O_2(X^1A) + 2\ H(^2S_{1/2})$$

(11a) $$[(H_2O(X^1A_1))_2] \rightarrow [H_2(X^1\Sigma_g^+)\ldots H_2O(X^1A_1)\ldots O(^1D)] \rightarrow H_2(X^1\Sigma_g^+) + H_2O_2(X^1A)$$

(11b) $$[(H_2O(X^1A_1))_2] \rightarrow [H_2(X^1\Sigma_g^+)\ldots H_2O(X^1A_1)\ldots O(^1D)] \rightarrow [H_2(X^1\Sigma_g^+)\ldots H_2OO(X^1A)] \rightarrow H_2(X^1\Sigma_g^+) + H_2O_2(X^1A)$$

Note that at 12 K, the water matrix still stores highly reactive radicals such as *thermalized* hydroxyl radicals, $OH(X^2\Pi_\Omega)$, which are immobile at low temperatures. Once the temperature increases, hydroxyl radicals can diffuse and – once they encounter a second hydroxyl radical – can recombine to form hydrogen peroxide. This reaction, however, is diffusion limited. Could there be an alternative reaction mechanism involving oxygen atoms? Recall that upon a unimolecular decomposition of the water molecules, the oxygen atoms are formed in the first excited, $^1D$, state. When the electron irradiation stops, the production of excited atoms ceases, too. In the gas phase, $O(^1D)$ has a life time of about 110 - 150 s (Bhardwaj & Haider 2002; Mohammed 1990); in the solid state, the life time is even shorter, and after a 3600 s equilibration time at 12 K, all oxygen atoms decay to their $^3P$ ground state. The reactivity of ground state atoms with water is different compared to the dynamics of the electronically excited counterparts via equation (11) (Ge, et al. 2005). Therefore, the enhanced column density of hydrogen peroxide at elevated temperatures could be explained by a diffusion controlled recombination of two hydroxyl radicals. We investigate now if our conclusions correlate with the mass spectrometric detection of the hydrogen peroxide molecule during the heating of the ice sample (Figure 7). It is important to stress that the exact pumping speed of the hydrogen peroxide molecule is not known (Appendix A). Therefore, the data compiled in Table 6 only present lower limits. The quantity of $H_2O_2$ as a function of electron fluence can be also fit with the functional form of equation (2) with $a = 1.2 \pm 0.4 \times 10^{15}$ molecules $cm^{-2}$ and $k = 8.1 \pm 1.2 \times 10^{-17}$ $electron^{-1}$ $cm^2$ (here, the '*t*' in the equation represents the electron fluence).

### 4.2. Formation of Atomic and Molecular Hydrogen

The data and mechanistical discussion on the formation of the hydrogen peroxide molecule also help us to understand the synthesis of atomic and molecular hydrogen during the irradiation of the ice samples. Here, equations (10) and (11) suggest that molecular hydrogen can be formed in a one-step mechanism via an electron induced decomposition of the water molecule (equation (11)). Alternatively the hydrogen atoms formed via equation (10) can recombine to form molecular hydrogen. At the present stage, our experimental data cannot quantify the branching ratio of an atomic versus molecular hydrogen production rate. However, the mass spectrometric detection of hydrogen atoms during the irradiation phase (section 3.2) is a direct proof that reaction (10) takes place. Likewise, the observation of oxygen atoms during the electron exposure suggests that reaction (11) is also an important pathway. Table 6 also shows that the water matrix stores hydrogen even when the electron irradiation is terminated. Here, $3.3 \pm 0.9 \times 10^{15}$ molecules (10 nA), $1.4 \pm 0.4 \times 10^{16}$ molecules (100 nA), $2.1 \pm 0.6 \times 10^{16}$ molecules (1000 nA), and $2.8 \pm 0.9 \times 10^{16}$ molecules (10000 nA) were released during the heating of the sample. Note that the evolution of the molecular hydrogen molecules upon the electron fluence can be also fit via equation (2) with $a = 1.5 \pm 0.5 \times 10^{16}$ molecules $cm^{-2}$ and $k = 8.1 \pm 1.2 \times 10^{-17}$ $electron^{-1}$ $cm^2$.

The trapping of molecular hydrogen inside water ice has been previously reported by Bar-Nun et al. (1987; 1988). Bar-Nun et al. (1985) also studied the ejection of water, molecular oxygen, as well as atomic and molecular hydrogen from water ice by 0.5 - 6 keV $H^+$ and $Ne^+$ ion bombardments. A review (Johnson & Quickenden 1997) of previous works likewise implied that hydrogen generated by irradiation processes might be ejected instead of being trapped inside water ice, therefore, it has been suggested that hydrogen peroxide and even molecular oxygen are more favorable reaction products which were proposed to be formed during the actual irradiation and the annealing. On the other hand, the experiment conducted by Sandford et al. (1993) suggested that significant amounts of $H_2$ may still be incorporated into the dust grains in molecular clouds through in situ production; Watanabe et al. (2000) also found that most of the $D_2$ molecules formed from UV photon irradiation of $D_2O$ ice were trapped inside the ice. Our experimental results are in agreement with those of Sandford and Watanabe. Here, most of the products produced by the electron irradiation at 12 K were trapped inside the water ice. In addition, we found that molecular hydrogen and hydrogen peroxide are the dominant products, while the production of molecular oxygen is smaller (Figure 7). These differences can be readily understood in terms of the energy transfer processes from the impinging charged particles (ions versus electrons) to the target molecules. SRIM (Ziegler 1992; Ziegler, Biersack, & Littmark 1985) and CASINO (Hovington, Drouin, & Gauvin 1997) calculations yield strong differences in the elastic stopping powers ($S_n$; the energy released via interaction of the implant with the target atoms via a screened Coulomb potential) and inelastic energy transfer processes ($S_e$). Table 5 compiles the nuclear and inelastic stopping powers. These data suggest that the release of the reaction products into the gas phase as reviewed by Johnson & Quickenden (1997) could result from an enhanced nuclear interaction potential which is simply absent in our electron irradiation experiments.

Table 5: Nuclear and inelastic stopping powers of various implants into water ice.

| irradiating particle | LET($S_e$), keV $\mu m^{-1}$ | LET($S_n$), keV $\mu m^{-1}$ |
|---|---|---|
| $e^-$, 5 keV | $4.3 \pm 0.1$ | - |
| $H^+$, 0.5 keV | $9.8 \pm 0.2$ | $4.0 \pm 0.4$ |
| $H^+$, 6.0 keV | $32.0 \pm 3.0$ | $1.1 \pm 0.1$ |
| $He^+$, 0.5 keV | $6.9 \pm 0.7$ | $22.7 \pm 2.0$ |
| $He^+$, 6.0 keV | $23.8 \pm 3.0$ | $10.8 \pm 1.0$ |

### 4.3. Formation of Atomic and Molecular Oxygen

In our vacuum system, the pumping speed of $H_2O_2$ is larger than that of $O_2$. Therefore, from Table 6 we can conclude that the quantity of molecular oxygen formed in the experiments is at least one order of magnitude lower than that the newly synthesized hydrogen peroxide molecules. Since $H_2$, $O_2$ and $H_2O_2$ originate from the reacting water molecules, the stoichiometry of a water molecule of H : O = 2 : 1 dictates the following balanced equation (12)).

(12) $$[2 \times N(H_2) + 2 \times N(H_2O_2)] / [2 \times N(H_2O_2) + 2 \times N(O_2)] = 2 : 1.$$

Here, the $[2 \times N(H_2) + 2 \times N(H_2O_2)]$ is the sum of the hydrogen atoms formed in molecular hydrogen and hydrogen peroxide, whereas the $[2 \times N(H_2O_2) + 2 \times N(O_2)]$ presents the number of oxygen atoms in the synthesized molecular oxygen and hydrogen peroxide molecules. This rearranges to equation (13).

(13) $$N(H_2) - N(H_2O_2) = 2 \times N(O_2).$$

Since $N(O_2) << N(H_2O_2)$, it can be easily seen that $N(H_2O_2) \approx N(H_2)$. However, a comparison of the total number of hydrogen molecules with the newly formed hydrogen peroxide (Table 6) molecules suggest that only about 10 – 15 % of the theoretically predicted hydrogen peroxide molecules have been detected. We would like to stress that that equations (12) and (13) neglect the hydroxyl radicals. Based on equation (5), hydroxyl radicals (OH) are certainly formed via a unimolecular decomposition of a water molecule upon electron irradiation. But, as stated in the previous section, *only* if two hydroxyl radicals are neighbors can they recombine to form a hydrogen peroxide molecule. The discrepancy of the theoretically predicted and actually observed hydrogen peroxide molecules could imply that the majority of the hydroxyl radicals (OH) are stored in the 12 K ice matrix. The infrared absorption of the hydroxyl radial at 3428 $cm^{-1}$ (Gerakines et al. 1996) overlaps with the $\nu_1$ and $\nu_3$ fundamental of the water molecule. This explains the failed detection of the hydroxyl radical in our experiments.

Finally, we would like to comment briefly on the possible formation mechanism of molecular oxygen in the ice samples. Sieger et al (Sieger, et al. 1998) proposed a two-step mechanism, in which molecular oxygen is formed via excitation and successive dissociation of a stable, but unidentified precursor molecule (possibly hydrogen peroxide). Also, several authors proposed a decomposition of the $HO_2(X^2A'')$ radical to molecular oxygen plus atomic hydrogen (equation (14); $\Delta_R G$ = 180.7 $kJmol^{-1}$). The latter was suggested to be formed either from recombination of thermalized oxygen atoms with hydroxyl radicals ($\Delta_R G$ = -243.3 $kJmol^{-1}$) (equation (15)) or by a charged particle induced fragmentation of the hydrogen peroxide molecule (equation (16); $\Delta_R G$ = 331.5 $kJmol^{-1}$). Finally, to generate the $HO_2$ radical inside water ice, another possible reaction is that hydrogen atoms react with molecular oxygen (the reverse reaction of equation (14)). However, this requires a large concentration of molecular oxygen in the water ice. Even so, there is no report on the formation of the $HO_2$ radical in the irradiation experiment of a water/molecular oxygen mixture (Moore & Hudson 2000). Instead, the authors observed the formation of ozone ($O_3$). As a matter of fact, the previous detection of the $HO_2$ radical by Gerakines et al (1996) is not well documented. First, the authors did not present an infrared spectrum of the absorptions claimed to be from the $HO_2$ radical. Most importantly, Gerakines et al's assignment of the 1389 $cm^{-1}$ band is incorrect; this absorption belongs to the hydrogen peroxide molecule (Table 4; asymmetric bending mode) and not to the $HO_2$ radical. Also, we would like to point out that the profiles of the $D_2O_2$ and $DO_2$ products as observed by Pan et al. (2004) are within the error limits virtually identical, and can be simply explained as $DO_2^+$ fragments from the $D_2O_2^+$ ion. Based on these considerations, we see no compelling experimental evidence for the involvement of the $DO_2/HO_2$ radical in the formation of hydrogen peroxide to date.

(14) $$HO_2(X^2A'') \rightarrow H(^2S) + O_2(X^3\Sigma_g^-)$$

(15) $OH(X^2\Pi) + O(^3P) \rightarrow HO_2(X^2A'')$

(16) $H_2O_2(X^1A) \rightarrow HO_2(X^2A'') + H(^2S)$

It is worth mention that we did observe a small amount of $HO_2^+$ at m/e = 33 with our mass spectrometer during the warm up of the irradiated samples. However, the patterns overlap exactly with the ion profile of singly ionized hydrogen peroxide; as a matter of fact, the intensity of $HO_2^+$ is about 3.4 % relative to $H_2O_2^+$. Therefore, $HO_2^+$ originates from the dissociation ionization of the hydrogen peroxide parent molecule $H_2O_2$ in the ionizer of the mass spectrometer. Therefore, we have no experimental evidence on the involvement of the $HO_2$ radical in the formation of water. Based on our results, we would like to discuss two feasible reaction mechanisms. Firstly, a recombination of two thermal oxygen atoms could form an oxygen molecule via equation (17). Alternatively, an oxywater molecule synthesized via equation (11b) might decompose to give molecular hydrogen plus molecular oxygen – initially in its first excited singlet state (equation (18)).

(17) $O(^3P) + O(^3P) \rightarrow O_2(X^3\Sigma_g^-)$

(18) $H_2OO(X^1A) \rightarrow O_2(a^1\Delta_g) + H_2(X^1\Sigma_g^+)$

The involvement of reaction (18) via (11b) could be rationalized in terms of the fit of the number of synthesized oxygen molecules versus the electron fluence via equation (2) suggesting a pseudo first order mechanism with $a = 0.91 \pm 0.3 \times 10^{14}$ molecules $cm^{-2}$ and $k = 8.1 \pm 1.2 \times 10^{-17}$ $electron^{-1}$ $cm^2$ (Figure 7). Future experiments utilizing partially isotope-labeled samples will help to pin down whether the reaction proceeds solely via equation (17) or (18) or involves both pathways. Note that plot of the newly synthesized hydrogen, oxygen, and hydrogen peroxide molecules versus the irradiation fluence could all be fit with an identical k value of $8.1 \pm 1.2 \times 10^{-17}$ $electron^{-1}$ $cm^2$.

Finally, we would like to investigate if the proposed reaction mechanisms can be supported energetically. This is done here for the irradiation experiment at 10 nA as a case study. During the irradiation, the sample was exposed to $5.7 \pm 0.2 \times 10^{10}$ electrons $s^{-1}$. Considering the ice thickness of $115 \pm 30$ nm and a linear energy transfer of $4.3 \pm 0.1$ keV $\mu m^{-1}$, $3.5 \pm 0.9 \times 10^{17}$ eV are absorbed within the ice sample after an irradiation time of 180 min. This corresponds to an average dose of $0.48 \pm 0.05$ eV per water molecule at a nominal electron current of 10 nA. In the worst case scenario - the limiting case that all the molecular hydrogen is formed via recombina-tion of hydrogen atoms released via equation (5) – 3.4 ± 0.6 × $10^{16}$ eV are necessary to account for the formation of the molecular hydrogen and hydrogen peroxide. This means that at small electron currents of 10 nA, 9.8 ± 1.5 % of the absorbed energy is being directed into a synthesis of newly formed molecules.

Table 6: Total number of synthesized oxygen, hydrogen, and hydrogen peroxide molecules during the electron irradiation experiments at various currents determined via mass spectrometry and FTIR spectroscopy. The area of irradiation was 1.86 ± 0.02 $cm^2$.

| | 10 nA | 100 nA | 1000 nA | 10000 nA |
|---|---|---|---|---|
| $O_2$ (warming) | >0 | >$3.11\times10^{13}$ | >$1.29\times10^{14}$ | >$1.71\times10^{14}$ |
| $O_2$ (irradiation) | 0 | 0 | >$8.41\times10^{13}$ | >$5.84\times10^{14}$ |
| $O_2$ (total) | 0 | >$3.11\times10^{13}$ | >$2.13\times10^{14}$ | >$7.55\times10^{14}$ |
| | | | | |
| $H_2O_2$ (warming) | >$6.83\times10^{13}$ | >$1.28\times10^{15}$ | >$1.85\times10^{15}$ | >$1.89\times10^{15}$ |
| $H_2O_2$ (irradiation) | 0 | 0 | 0 | >$9.17\times10^{13}$ |
| $H_2O_2$ (total) | >$6.83\times10^{13}$ | >$1.28\times10^{15}$ | >$1.85\times10^{15}$ | >$1.98\times10^{15}$ |
| | | | | |
| $H_2$ (warming) | $3.30\times10^{15}$ | $1.35\times10^{16}$ | $2.10\times10^{16}$ | $2.79\times10^{16}$ |
| $H_2$ (irradiation) | 0 | 0 | $1.82\times10^{15}$ | $1.43\times10^{16}$ |
| $H_2$ (total) | $3.30\times10^{15}$ | $1.35\times10^{16}$ | $2.28\times10^{16}$ | $4.22\times10^{16}$ |
| | | | | |
| $H_2O_2$ (FTIR) | 0 | $1.5\times10^{15}$ | $4.3\times10^{15}$ | $1.1\times10^{16}$ |

## 5. Astrophysical Implications

The experimental results on the electron irradiation of crystalline water ice hold wide reaching implications for astrochemistry and also for astrobiology. First, water ice serves as the energy-transfer medium and active participant in a variety of radiation-driven organic chemistry reactions thought to be important in either the molecular clouds or the primordial solar nebula (Dressler 2001; Ehrenfreund, et al. 1999; Minh & Dishoeck 2000; Sykes 2002). Here, most of the organic molecules are thought to be formed inside water-rich matrices. On the other hand, we demonstrated that a radiation induced processing of water can also generate – besides molecular oxygen and hydrogen peroxide - highly reactive, suprathermal and/or electronically excited atoms such as hydrogen and oxygen). These atoms can subsequently react with newly formed astrobiologically important molecules even inside ices and effectively degrade them. The reactivity of suprathermal atoms with double and triple bonds in organic molecules is well known (Kaiser 2002) and actually complicated the formation of astrobiologically significant building blocks such as aminoacids in water-rich ices. Also, hydrogen peroxide itself stores highly reactive oxygen atoms and hydroxyl radicals; here, upon interaction of ionizing radiation, organic molecules neighboring to a hydrogen peroxide molecule can be degraded easily. Therefore, before we can fully understand those processes participated in by water, we need to have enough knowledge about water itself. The systematic study of the effects of irradiation on water ice as presented here have therefore strong implications on future studies of the stability of organic molecules in water–rich matrices as present in the cold molecular clouds, Kuiper Belt Objects, and Oort cloud.

Secondly, the production rates and kinetic equations can help to predict quantitatively the conversion of water into molecular hydrogen, oxygen, and hydrogen peroxide on low temperature ices in cold molecular clouds (Ehrenfreund, et al. 1999; Sykes 2002) and on comets stored in Oorts cloud. In our experiments, the water ice irradiated reaches, for example, an equilibrium concentration of hydrogen peroxide after a dose of about 19 eV per water molecule; this converted about 0.6 % of the water into hydrogen peroxide. Therefore, comets in Oorts cloud will reach an equilibrium concentration of hydrogen peroxide molecules after about $0.2 \times 10^6$ years at 10 K and after about $0.1 \times 10^6$ years for Kuiper Belt Objects (Yeghikyan & Kaiser 2005). However, we would like to stress that the situation in Kuiper Belt Objects is more complex since thermal processes can also take part at 40 – 50 K. We notice that Moore & Hudson (Moore & Hudson 2000) studied the irradiation of pure water ice and mixtures with molecular oxygen at 80 K. They observed the 3.5μm absorption feature of $H_2O_2$ in the $O_2/H_2O$ mixture, but the production of $H_2O_2$ was lower than the detectable level in the pure water ice. That brings out some questions. Does the existence of $O_2$ in the surface of icy satellites help to produce more $H_2O_2$, or does the production of $H_2O_2$ help to generate more $O_2$ in icy satellites? Which one comes first? Therefore, absolute production rates of newly formed molecules in water-irradiated ices are also required at higher temperatures; these studies are currently in progress in our laboratory.

Thirdly, we were able to quantify the ratio of molecular oxygen versus hydrogen peroxide formed in the electron exposure of water ices at 10 K. Since hydrogen peroxide has strong infrared absorptions but molecular oxygen does not, our study may help to calculate – once the hydrogen peroxide column densities in outer solar system objects or on interstellar grains are measured –the molecular oxygen abundance in these ices. In our irradiation experiments, the production of molecular oxygen is at least one magnitude lower than that of hydrogen peroxide. Therefore, we suggest that the concentration of molecular oxygen in the dust grains in molecular clouds and Oorts cloud objects should be lower than the concentration of hydrogen peroxide if those two species are formed from irradiation of pure water ice alone.

Fourthly, our experiments demonstrated clearly that most of the molecular hydrogen generated in water ice at 12K is not released until the ice is warmed up to 100 – 140 K. Therefore, comets in the Oort cloud are predicted to store a large amount of molecular hydrogen gas inside water ice, and a strong molecular hydrogen coma might be formed when those comets approach to ~ 10 AU distance from the sun. The possible formation of hydrogen comae around comets at large heliocentric distances has also been suggested by Bar-Nun et al (Bar-Nun & Prialnik 1988). However, it needs to be pointed out that all laboratory experiments have very different time scales from the astronomical objects. The laboratory experimental timescale is of the order of hours, whereas the time scale for the objects in our solar system is millions to billions of years. Does the molecular hydrogen actually "leak out'' of comets at a low rate at low temperatures (10 – 50 K), which is negligible in the laboratory but significant in space?

Finally, our experiments show that molecular oxygen can be trapped inside crystalline water ice at temperature as high as 147 K. The temperatures on the surface of Jovian satellites such as Ganymede, Europa, and Callisto are lower than that. So we are expecting some molecular oxygen to be trapped inside the ice in the surfaces of those icy satellites. That is consistent with astronomical observations (Spencer, et al. 1995; Spencer & Calvin 2002). Most importantly, however, our studies suggest that not even trace amounts of ozone can be formed in pure water samples during electron irradiation processes. This can be the result of a pure electronic interaction of the implant with the water molecules and hence an inherent lower production rate of oxygen atoms compared to direct knock-on processes induced by solar wind particles; alter-natively, the enhanced concentration of hydrogen atoms in the target might prevent a significant concentration of free oxygen atoms reaction to ozone. This will be investigated also in future laboratory experiments.

## 6. Summary

We conducted a systematic study of the irradiation of cubic crystalline water ice in an ultra-high vacuum machine. Reaction mechanisms to synthesize experimentally observed hydrogen and oxygen both in its atomic (H, O) and molecular ($H_2$, $O_2$) form together with hydrogen peroxide ($H_2O_2$) are discussed. Additional formation routes were derived from the sublimation profiles of molecular hydrogen (90 – 140 K), molecular oxygen (147 – 151 K) and hydrogen peroxide (170 K). We also presented quantitative evidence on the involvement of hydroxyl radicals and possibly oxygen atoms as building blocks to yield hydrogen peroxide at low temperatures (12 K) and via a diffusion-controlled mechanism in the warming up phase of the irradiated sample. These studies are a first step in a systematic understanding of the charged particle processing of water ices in the molecular clouds and in outer solar system objects, such as Kuiper Belt Objects, Oort cloud, comets, and icy satellites. Further experiments are planned to resolve the temperature dependent formation routes of H, O, $H_2$, $O_2$, and $H_2O_2$ in water ice and also to quantify potential differences in irradiation processes in crystalline versus amorphous water ices.

## Acknowledgements

This work was supported by the NASA Astrobiology Institute under Cooperative Agreement NNA04CC08A at the Institute for Astronomy at the University of Hawaii-Manoa (WZ, DJ, RIK). We are also grateful to Ed Kawamura (University of Hawaii at Manoa, Department of Chemistry) for his electrical work.

## Appendix

The temporal evolution of the ion currents of distinct m/e ratios help to compute the partial pressures of the gases. Here, the partial pressure of a gas j, $p_j$, is proportional to the ion current of the j-th species at a i-th mass-to-charge m/e, $I_{j,i}$, equation (A1). Therefore, if we know the ion current and the proportionality constant we can then compute the partial pressure (Chambers, et al. 1998); if fragments of the neutral species have the same mass-to-charge rations, matrix interval arithmetic has to be utilized (Kaiser et al. 1995).

(A1) $$I_{j,i} = p_j \times f_{j,i}$$

To extract the proportionality constants, the calibration experiments were conducted in separate experiments by leaking the pure products gases ($H_2$, $O_2$, $H_2O$) into the main chamber. The intensities of the mass fragments were recorded in dependence on the different inlet pressures, i.e. different pressure readings of the ion gauge (Granville Phillips; extractor type gauge). We have to account now for the relative sensitivities of the ion gauge $R_j$ with respect to the inlet gas via equation (A2).

(A2) $$p_{j(real)} = p_{j(reading)} \times R_j$$

By plotting the ion current $I_{j,i}$ versus the partial pressure $p_{j(real)}$, the proportionality constant $f_{j,i}$ can be obtained. This procedure yields $f_{H2,2}$ = 3.1, $f_{O2,32}$ = 1.5, and $f_{H2O,18}$ = 0.9. Since the calibration with neat $H_2O_2$ is not feasible, we estimated the conversion factor of $H_2O_2$ to be about 1.0. Note that the uncertainty of the relative ionization gauge sensitivities is ± 10 %. Due to geometric variations in electrode structures, the ion gauge is accurate within ± 20 %. We estimate the error bar of the partial pressure is ± 30 %.

With the partial pressures known, we can calculate in principle the numbers of synthesized molecules. This requires the knowledge of the effective pumping speeds for each gas. The effective pumping speed, $S_{effective}$, can be computed via equation (3) from the pumping speed of the vacuum pump, S, and the conductance between the pump and the vacuum chamber, C.

($A$3) $$\frac{1}{S_{effective}} = \frac{1}{S} + \frac{1}{C}$$

In our experiments, it is hard to calculate the exact effective pumping speeds for oxygen and hydrogen peroxide, simply because they can also be condensed on the second stage of the cold head. Considering only the conductance of the vacuum chamber and the pumping of the turbo molecular pump, we calculated the pumping speed for $H_2$, $O_2$ and $H_2O_2$ to be 604 $ls^{-1}$, 686 $ls^{-1}$, and 686 $ls^{-1}$, respectively. Note that the pumping speeds for $O_2$ and $H_2O_2$ are minimum values since the pumping speed of the cold head is unknown and can not included. Therefore, only the pumping speed of $H_2$ can be calculated correctly. The uncertainty for the effective pumping speed of $H_2$ is ± 15 % (Chambers, et al. 1998). Now, the quantities of produced molecules can be computed via equation (A4) (Kaiser, et al. 1995). Here, $N_j$ is the number of molecules for species j, $N_A$ is Avogadro constant, $R$ is the molar gas constant, $T$ the temperature of the residual gas (here: 298 K), $p_j$ is the partial pressure of species j, and $S_{effective}$ is the effective pumping speed of species j.

$$(A4) \qquad N_j = \frac{\int_{t_1}^{t_2} p_j \times S_{effective}\, dt}{RT} \times N_A$$

Averna, D., & Pirronello, V. 1991, Astronomy and Astrophysics, 245, 239

Bahr, D. A., Fama, M., Vidal, R. A., & Baragiola, R. A. 2001, Journal Of Geophysical Research-Planets, 106, 33285

Bar-Nun, A., Dror, J., Kochavi, E., & Laufer, D. 1987, Physical Review B, 35, 2427

Bar-Nun, A., Herman, G., Rappaport, M. L., & Mekler, Y. 1985, Surface Science, 150, 143

Bar-Nun, A., Kleinfeld, I., & Kochavi, E. 1988, Physical Review B, 38, 7749

Bar-Nun, A., & Prialnik, D. 1988, Astrophysical Journal, 324, L31

Baragiola, R. L. A., Loeffler, M. J., Raut, U., Vidal, R. A., & Wilson, C. D. 2005, Radiation Physics And Chemistry, 72, 187

Bennett, C. J., Jamieson, C., Mebel, A. M., & Kaiser, R. I. 2004, Physical Chemistry Chemical Physics, 6, 735

Bhardwaj, A., & Haider, S. A. 2002, Advances in Space Research, 29, 745

Brown, W. L., et al. 1980, Nuclear Instruments & Methods, 170, 321

Brown, W. L., Augustyniak, W. M., & Lanzerotti, L. J. 1980, Physical Review Letters, 45, 1632

Brown, W. L., Lanzerotti, L. J., Poate, J. M., & Augustyniak, W. M. 1978, Physical Review Letters, 40, 1027

Campins, H., Swindle, T. D., & Kring, D. A. 2004, Cellular Origin and Life in Extreme Habitats and Astrobiology (2004), 6 (Origins: Genesis, Evoluation and Diversity of Life), 569-591.

Carlson, R. W., et al. 1999, Science, 283, 2062

Chambers, A., Fitch, R. K., & Halliday, B. S. 1998, Basic vacuum technology (2nd ed.; Bristol; Philadelphia: Institute of Physics Pub.)

Christiansen, J. W., Carpini, D. D., & Tsong, I. S. T. 1986, Nuclear Instruments & Methods In Physics Research Section B-Beam Interactions With Materials And Atoms, 15, 218

Cooper, B. H., & Tombrello, T. A. 1984, Radiation Effects And Defects In Solids, 80, 203

de Bergh, C. 2004, in ASSL Vol. 305: Astrobiology: Future Perspectives, 205

De Pater, I., & Lissauer, J. J. 2001, Planetary sciences (Cambridge; New York: Cambridge University Press)

Dressler, R. A. 2001, Chemical dynamics in extreme environments (Singapore; River Edge, NJ: World Scientific)

Ehrenfreund, P., Krafft, C., Kochan, H., & Pirronello, V. 1999, Laboratory astrophysics and space research (Boston: Kluwer Academic Publishers)

Flower, D. R., & Pineau-Des-Forets, G. 1990, Monthly Notices of the Royal Astronomical Society, 247, 500

Ge, Y. B., Head, J. D., & Kaiser, R. I. 2005, in preparation.

Gerakines, P. A., Schutte, W. A., & Ehrenfreund, P. 1996, Astronomy And Astrophysics, 312, 289

Giguere, P. A., & Harvey, K. B. 1959, Journal of Molecular Spectroscopy, 3, 36

Gomis, O., Leto, G., & Strazzulla, G. 2004a, Astronomy & Astrophysics, 420, 405

Gomis, O., Satorre, M. A., Strazzulla, G., & Leto, G. 2004b, Planetary And Space Science, 52, 371

Greenberg, J. M., van de Bult, C. E. P. M., & Allamandola, L. J. 1983, Journal of Physical Chemistry, 87, 4243

Hagen, W., Tielens, A., & Greenberg, J. M. 1981, Chemical Physics, 56, 367

Hall, D. T., Feldman, P. D., McGrath, M. A., & Strobel, D. F. 1998, Astrophysical Journal, 499, 475

Hall, D. T., Strobel, D. F., Feldman, P. D., McGrath, M. A., & Weaver, H. A. 1995, Nature, 373, 677

Hendrix, A. R., Barth, C. A., Stewart, A. I. F., Hord, C. W., & Lane, A. L. 1999. in Lunar and Planetary Institute Conference Abstracts, Hydrogen Peroxide on the Icy Galilean Satellites, 2043

Hornekaer, L., Baurichter, A., Petrunin, V. V., Field, D., & Luntz, A. C. 2003, Science, 302, 1943

Hornekaer, L., Baurichter, A., Petrunin, V. V., Luntz, A. C., Kay, B. D., & Al-Halabi, A. 2005, Journal of Chemical Physics, 122, 124701

Hovington, P., Drouin, D., & Gauvin, R. 1997, Scanning, 19, 1

Jamieson, C. S., Bennett, C. J., Mebel, A. M., & Kaiser, R. I. 2005, Astrophysical Journal, 624, 436

Jenniskens, P., Banham, S. F., Blake, D. F., & McCoustra, M. R. S. 1997, Journal Of Chemical Physics, 107, 1232

Jenniskens, P., Blake, D. F., & Kouchi, A. 1998. in ASSL Vol. 227: Solar System Ices, Amorphous Water Ice. a Solar System Material, 139

Jewitt, D. C., & Luu, J. 2004, Nature, 432, 731

Johnson, R. E. 1990, Energetic charged-particle interactions with atmospheres and surfaces (Berlin; New York: Springer-Verlag)

Johnson, R. E., & Quickenden, T. I. 1997, Journal Of Geophysical Research-Planets, 102, 10985

Kaiser, R. I. 2002, Chemical Reviews, 102, 1309

Kaiser, R. I., Jansen, P., Petersen, K., & Roessler, K. 1995, Review Of Scientific Instruments, 66, 5226

Kimmel, G. A., & Orlando, T. M. 1996, Physical Review Letters, 77, 3983

Kimmel, G. A., & Orlando, T. M. 1995, Physical Review Letters, 75, 2606

Kimmel, G. A., Orlando, T. M., Vezina, C., & Sanche, L. 1994, Journal of Chemical Physics, 101, 3282

Kouchi, A., & Kuroda, T. 1990, Nature, 344, 134

Liu, K. 2001, Annual Review of Physical Chemistry, 52, 139

Manico, G., Ragun, G., Pirronello, V., Roser, J. E., & Vidali, G. 2001, Astrophysical Journal, 548, L253

Matich, A. J., Bakker, M. G., Lennon, D., Quickenden, T. I., & Freeman, C. G. 1993, Journal Of Physical Chemistry, 97, 10539

Minh, Y. C., & Dishoeck, E. F. v. 2000, Astrochemistry, from molecular clouds to planetary systems: proceedings of the 197th Symposium of the International Astronomical Union held in Sogwipo, Cheju, Korea, 23-27 August 1999 (Provo, UT: Astronomical Society of the Pacific)

Mohammed, H. H. 1990, Journal Of Chemical Physics, 93, 412

Moore, M. H., & Hudson, R. L. 2000, Icarus, 145, 282

Noll, K. S., Roush, T. L., Cruikshank, D. P., Johnson, R. E., & Pendleton, Y. J. 1997, Nature, 388, 45

Orlando, T. M., & Sieger, M. T. 2003, Surface Science, 528, 1

Pan, X. N., Bass, A. D., Jay-Gerin, J. P., & Sanche, L. 2004, Icarus, 172, 521

Pirronello, V., & Averna, D. 1988, Astronomy and Astrophysics, 196, 201

Roser, J. E., Manico, G., Pirronello, V., & Vidali, G. 2002, Astrophysical Journal, 581, 276

Sandford, S. A., & Allamandola, L. J. 1993, Astrophysical Journal, 409, L65

Shematovich, V. I., Johnson, R. E., Cooper, J. F., & Wong, M. C. 2005, Icarus, 173, 480

Shi, M., Baragiola, R. A., Grosjean, D. E., Johnson, R. E., Jurac, S., & Schou, J. 1995a, Journal of Geophysical Research, 100, 26387

Shi, M., Grosjean, D. E., Schou, J., & Baragiola, R. A. 1995b, Nuclear Instruments & Methods In Physics Research Section B-Beam Interactions With Materials And Atoms, 96, 524

Sieger, M. T., Simpson, W. C., & Orlando, T. M. 1998, Nature, 394, 554

Spencer, J. R., & Calvin, W. M. 2002, Astronomical Journal, 124, 3400

Spencer, J. R., Calvin, W. M., & Person, M. J. 1995, J. Geophys. Res., 100, 19049

Sykes, M. V. 2002, The future of solar system exploration, 2003-2013: community contributions to the NRC solar system exploration decadal survey (1st ed.; San Francisco, CA: Astronomical Society of the Pacific)

Taylor, S. R. 2001, Solar system evolution: a new perspective: an inquiry into the chemical composition, origin, and evolution of the solar system (2nd ed.; Cambridge; New York: Cambridge University Press)

Tielens, A. G. G. M., & Hagen, W. 1982, Astronomy and Astrophysics, 114, 245

Vidal, R. A., Bahr, D., Baragiola, R. A., & Peters, M. 1997, Science, 276, 1839

Watanabe, N., Horii, T., & Kouchi, A. 2000, Astrophysical Journal, 541, 772

Westley, M. S., Baragiola, R. A., Johnson, R. E., & Baratta, G. A. 1995, Planetary and Space Science, 43, 1311

Yeghikyan, A., & Kaiser, R. I. 2005, in preparation.

Ziegler, J. F. 1992, Handbook of ion implantation technology (Amsterdam; New York: North-Holland)

Ziegler, J. F., Biersack, J. P., & Littmark, U. 1985, The stopping and range of ions in solids (New York: Pergamon)

**Figure Captions**

Fig. 1 Top view of the experimental setup.

Fig. 2 Infrared spectra of cubic crystalline water taken before irradiation (solid line) and after irradiation (dash line) at 12K. The insert enlarges the hydrogen peroxide absorption band centered at 2851 $cm^{-1}$ (3.5 μm).

Fig. 3 Infrared spectrum of the irradiated sample taken in the warm up phase at 170 K; most of the water has sublimed. Peaks A, B and C correspond to the $v_2+v_6$ combination mode，$v_2$ symmetric bending， and $v_6$ asymmetric bending of the hydrogen peroxide molecule. The remaining absorption features belong to water ice.

Fig. 4 The temporal evolution of the $H_2O_2$ column density during irradiation exposure at electron currents of 100 nA (a), 1000 nA (b), and 10000 nA (c). The red lines present the best fits (see text for a detailed discussion).

Fig. 5 The change of the partial pressure of molecular hydrogen as a function of the irradiation time for electron currents of 0, 10, 100, 1000, and 10000 nA.

Fig. 6 The temporal evolution of the ion currents of molecular hydrogen ($H_2$), molecular oxygen ($O_2$), hydrogen peroxide ($H_2O_2$), and water ($H_2O$) during the warm up phase experiment. (a) 0 nA, (b) 10nA, (c) 100nA, (d) 1,000nA, and (e) 10,000nA

Fig. 7 Quantification of the numbers of molecular hydrogen ($H_2$) (a), molecular oxygen ($O_2$) (b), and hydrogen peroxide ($H_2O_2$) (c) as a function of the irradiation electron fluence.

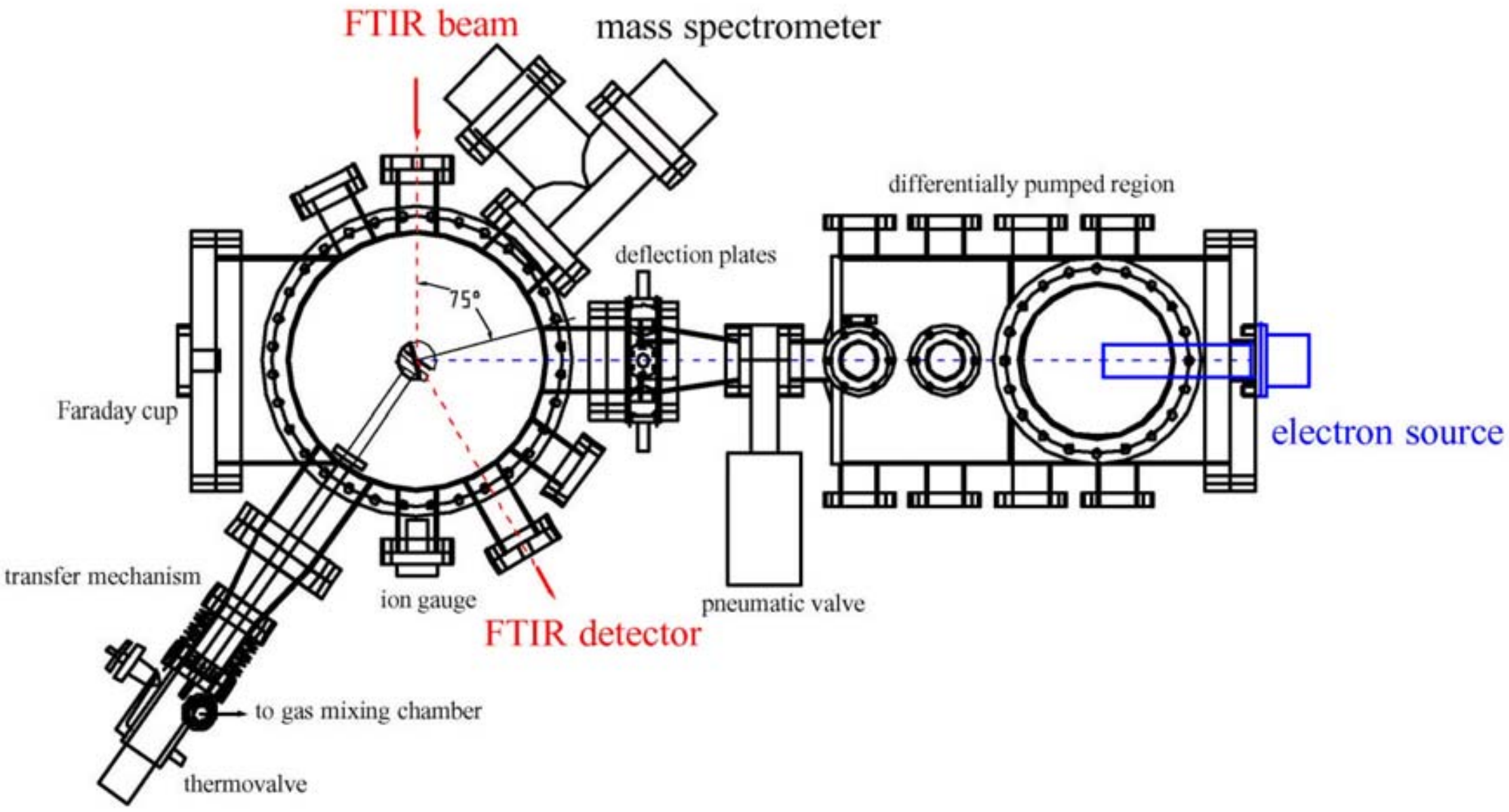

Fig. 1

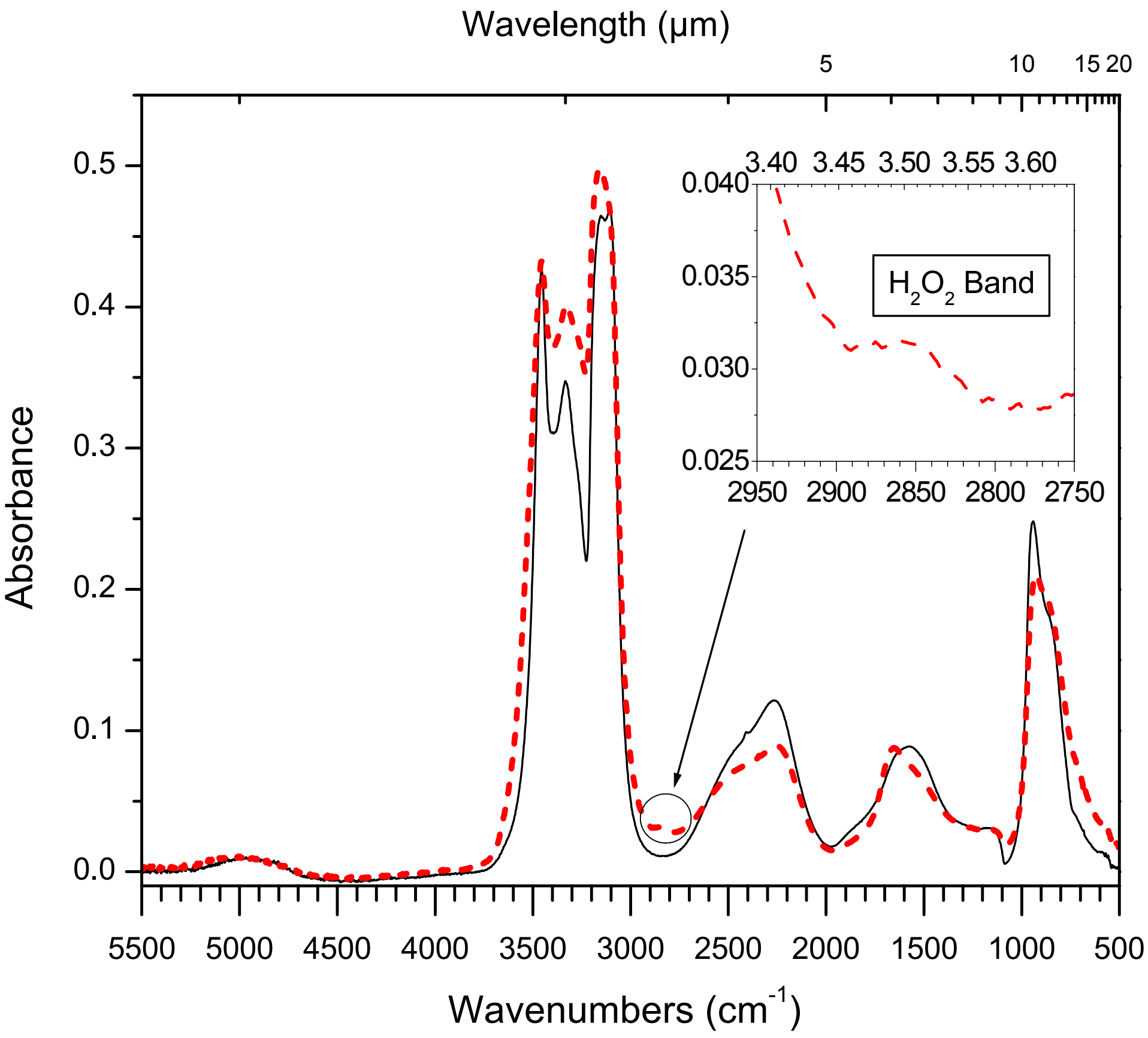
Wavelength (μm)
5
10
15 20
0.5
0.4
0.3
0.2
0.1
0.0
Absorbance
3.40 3.45 3.50 3.55 3.60
0.040
0.035
0.030
0.025
2950 2900 2850 2800 2750
H2O2 Band
5500 5000 4500 4000 3500 3000 2500 2000 1500 1000 500
Wavenumbers (cm-1)

Fig. 2

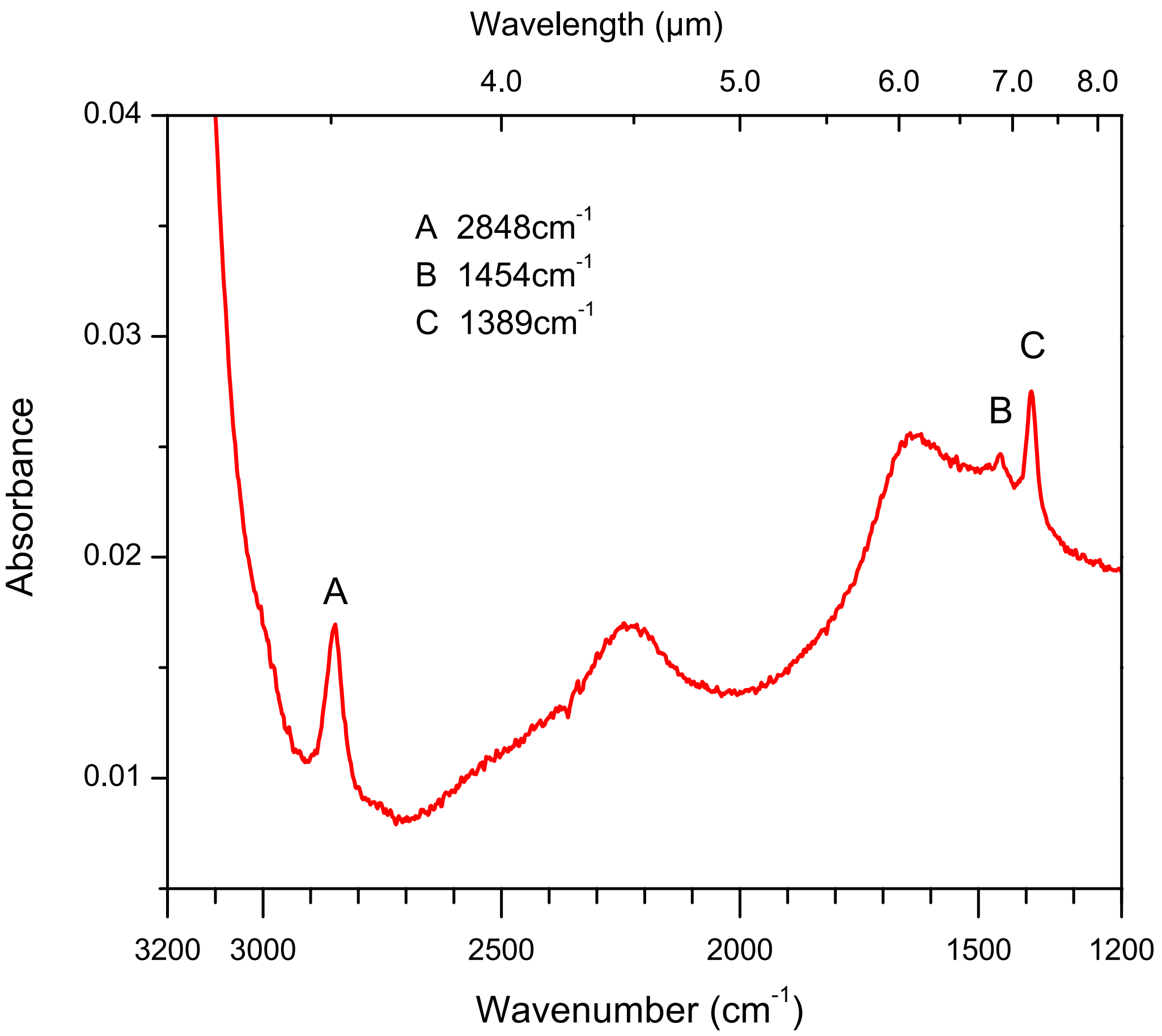

Fig. 3

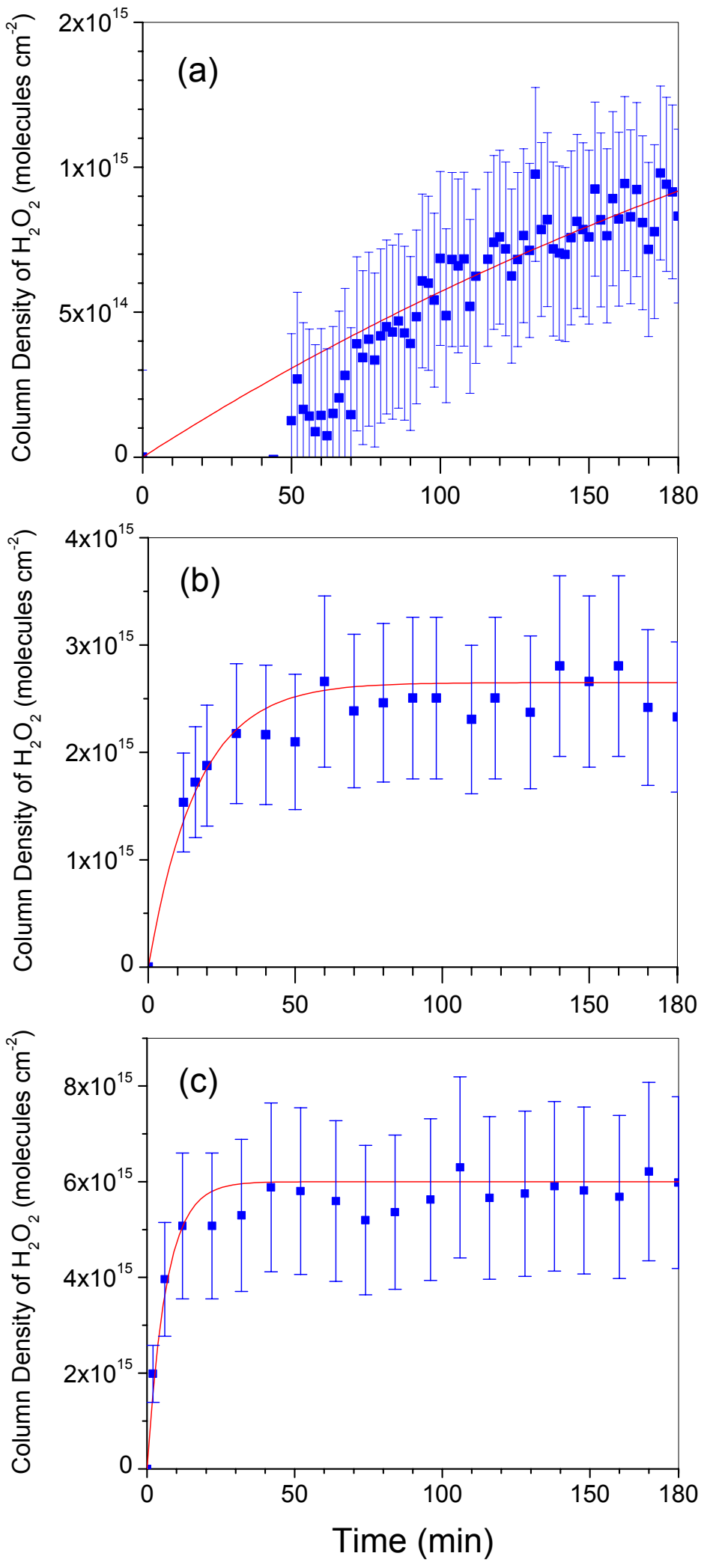

Fig. 4

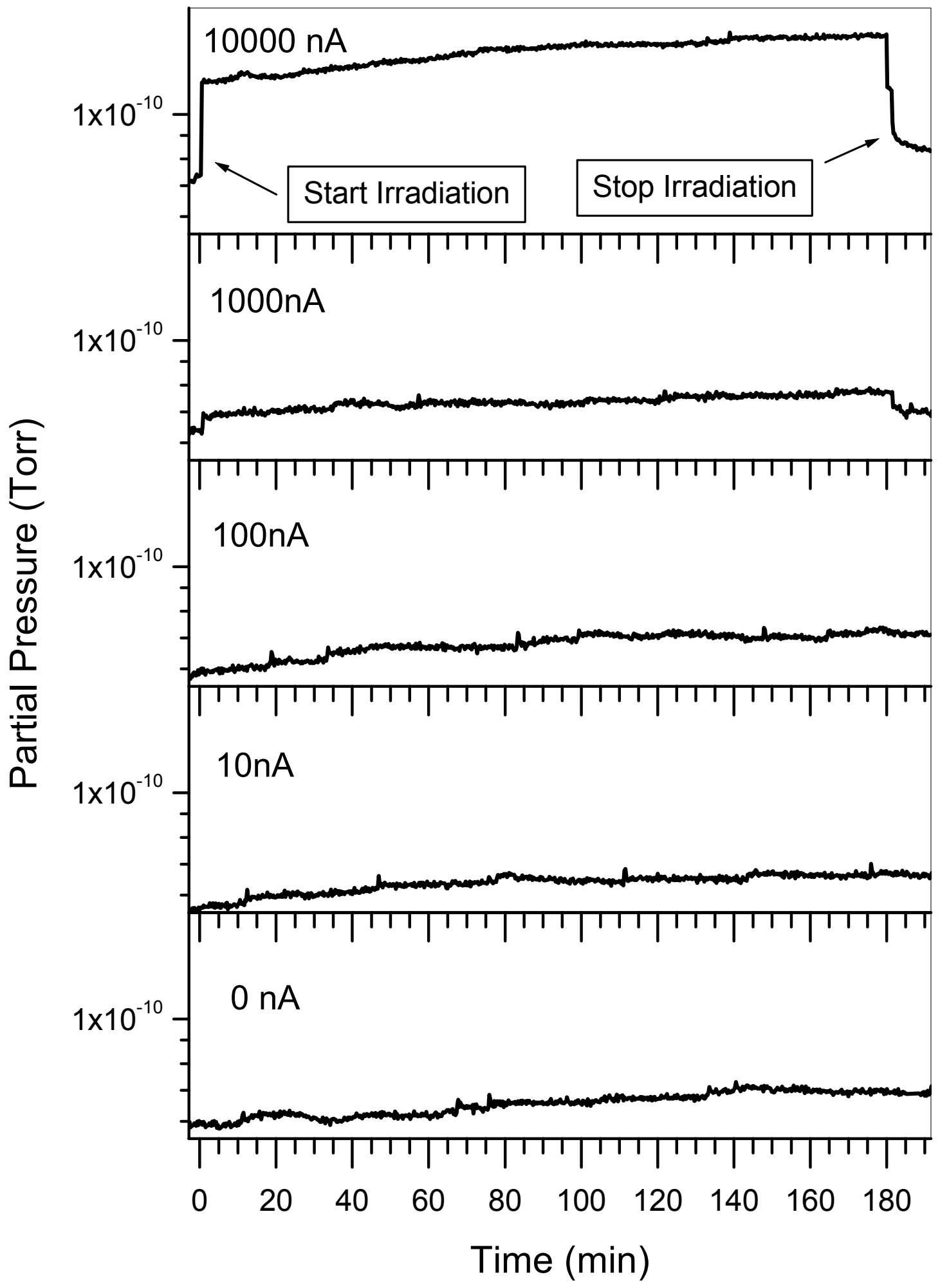

Fig. 5

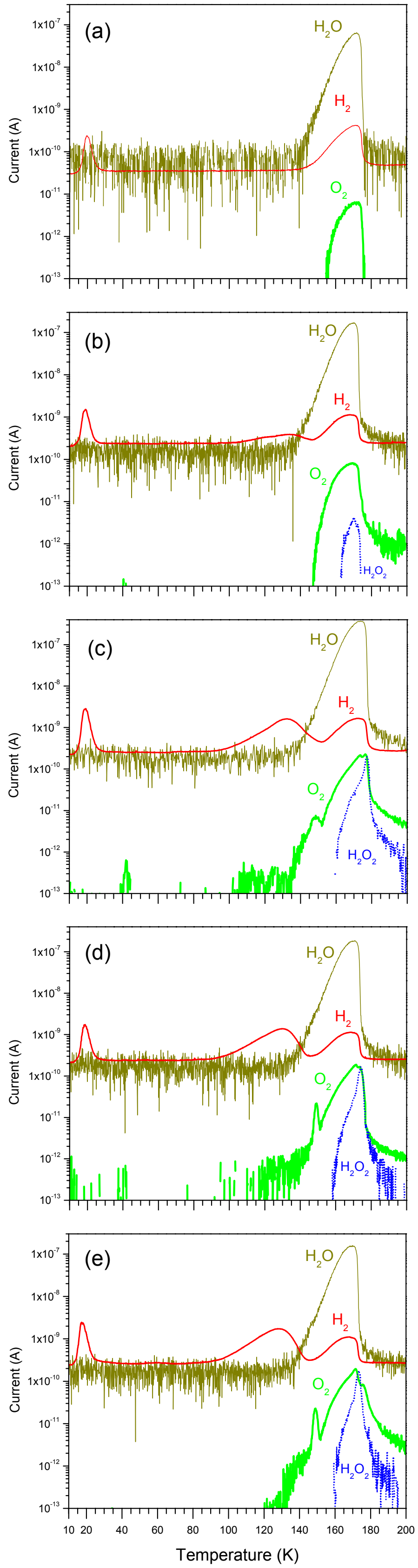

Fig. 6

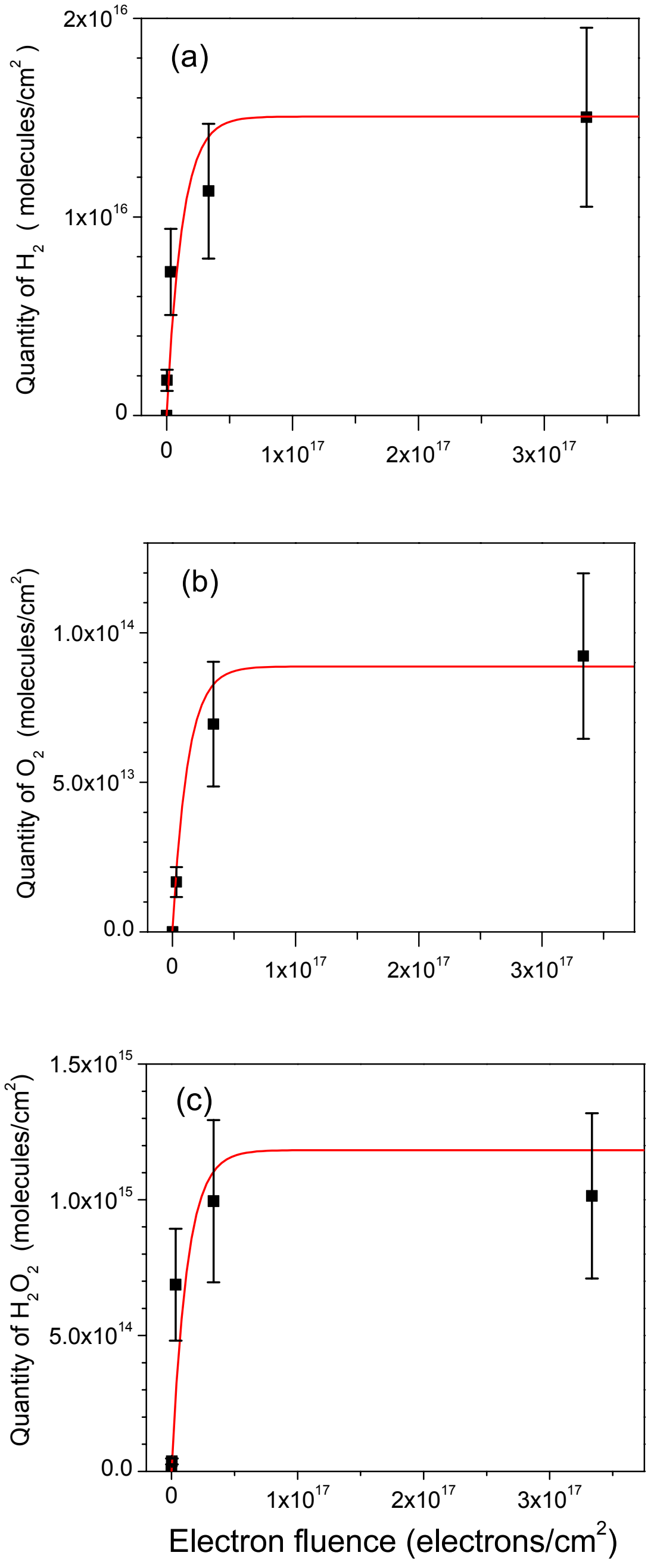

Fig. 7